\documentstyle[12pt,apjpt4]{article}
\input{psfig}

\begin{document}
\begin{titlepage}

\title{ \huge New Insights from HST Studies of Globular Cluster Systems II:  Analysis of 29 S0 Systems \altaffilmark{1}    }
\vspace*{0.5 truein}

\author{ Arunav Kundu \altaffilmark{2} }
\vspace*{0.2 truein}
\affil{ Space Telescope Science Institute, 3700 San Martin Drive, Baltimore, MD 21218 }
\vspace*{0.2 truein}
\affil{  Electronic Mail: akundu@astro.yale.edu}
\vspace*{0.2 truein}
\affil{ and }
\vspace*{0.2 truein}
\affil{ Dept of Astronomy, University of Maryland, College Park, MD 20742-2421 }
\vspace*{0.5 truein}

\author{ Bradley C. Whitmore }
\vspace*{0.2 truein}
\affil{ Space Telescope Science Institute, 3700 San Martin Drive, Baltimore, MD 21218 }
\vspace*{0.2 truein}
\affil{  Electronic Mail: whitmore@stsci.edu }

\altaffiltext{1}{\normalsize Based on observations with the 
NASA/ESA {\it Hubble Space Telescope}, obtained at the Space Telescope Science 
Institute, which is operated by the Association of Universities for Research 
in Astronomy, Inc., under NASA contract NAS5-26555 }

\altaffiltext{1}{\normalsize Present address:  Astronomy Department, Yale University, 260 Whitney Av., New Haven, CT 06511} 

 \begin{abstract}
\end{abstract}
	We examine the globular cluster systems (GCS) of a sample of 34 S0 galaxies from a WFPC2 snapshot survey in the V and I-bands. Of these 34 galaxies, 29 have
 measurable globular cluster systems.  The mean color of the GCSs of individual galaxies 
vary from V-I=0.85 mag  to V-I=1.17 mag.
The average color of GCSs in all 29 S0 galaxies, V-I=1.00$\pm$0.07 mag, is similar to the value of V-I=1.04$\pm$0.04 derived for ellipticals in a companion paper. The mean metallicity of a cluster system is correlated to the
luminosity (or mass) of the host galaxy, but is not dependent on the Hubble type. Our measurements of the local specific frequency, on the other hand, confirm that the cluster formation efficiency is a function of Hubble type. The mean local specific frequency of our sample within the WFPC2 field-of-view is 1.0$\pm$0.6, lower than
 S$_{N(Local)}$=2.4$\pm$1.8 derived for ellipticals in a similar analysis.
 Although we are able to confirm a bimodal color distribution in only one galaxy from the shallow 'snapshot' images analyzed in this paper statistical tests suggest that 10-20$\%$ of S0s are bimodal {\it at the present level of photometric accuracy}.  There are no significant trends in GCS properties with membership or location of the S0 host in a galaxy cluster. We have measured the turnover luminosity of the globular cluster luminosity function (GCLF) for the richest
few GCSs and find the GCLF distances to be in agreement with other estimates.  The  globular clusters in S0 galaxies have average half-light radii of $\approx$2.6 pc which is similar to that of other galaxies, including our own.

\end{titlepage}
\section{Introduction}
	Globular clusters (GC) are among the most pristine objects in galaxies and hence  provide
 invaluable insights into the chemical and kinematic conditions prevalent
 during the early epochs of galaxy formation. Just how early is presently 
a matter of some debate. Initial studies of the globular cluster systems
of the Milky Way and other nearby galaxies led to the conclusion
 that all globular
clusters are old, metal poor systems that were formed during
the initial collapse phase of a protogalactic gas cloud. The subsequent 
discovery of metal rich GCSs and bimodal color (metallicity) distributions in some 
galaxies has led to much discussion about alternate mechanisms of globular cluster 
formation.  Schweizer (1987) and Ashman $\&$
 Zepf (1992) proposed that some GCs may be formed during the interaction or merger
of galaxies. While Zepf $\&$ Ashman (1993) suggest that the bimodality in the color
distribution is a natural consequence of the merger scenario, Forbes, Brodie \&
Grillmair (1997) favor multiple phases of cluster formation during the 
collapse of a gas cloud into a galaxy.

	 Only recently has it become possible to test these and other  theories of 
globular cluster formation and evolution, due to the explosion in the number 
of observations of GCSs of external galaxies. However, much of the attention has
been focused on cluster-rich ellipticals, with S0s being highly
 underrepresented  in the sample of globular cluster systems studied to date.
 This is rather unfortunate since S0s may provide some essential
clues to explain the small systematic differences between the GCSs of
 ellipticals and spirals, and may hold some of the keys to understanding 
the galaxy formation process. In this paper we attempt to 
 rectify this imbalance
by analyzing the  cluster systems of 34 S0 galaxies from archival Hubble Space Telescope (HST) images. In a companion paper, Kundu \& Whitmore (2001)  (hereafter referred to as Paper I), we present a similar study of elliptical systems.

\section{Observations and Data Reduction}

	 This study is based on  archival HST Wide
Field and Planetary Camera 2 (WFPC2; hereafter WF=Wide Field Camera and PC=Planetary 
Camera) short exposure, 'snapshot' images of nearby S0 galaxies. The list of program galaxies along with 
some salient information is presented  in Table 1 (Note: Throughout this paper
tables and figures are sorted by absolute magnitude of the host galaxy from Table 1, except in cases where distance-dependent effects are important, where they are sorted by distance).  The galaxies were
 imaged with the F555W (1 image$\times$160s each) and F814W (2 images $\times$ 
160s each) filters  between Sep 6 1995 and May 3 1996.

  We used the STSDAS utility GCOMBINE and the IRAF task COSMICRAYS  to remove
 cosmic rays and hot pixels respectively from the F814W images, while we used only
 COSMICRAYS to reject cosmic ray events from the single F555W snapshots.
 We then identified  cluster candidates 
using the technique described in detail in Kundu \& Whitmore (1998). Briefly, DAOPHOT is used
to identify sources using a low cutoff (1.5$\sigma$, where $\sigma$ is the 
standard deviation in the smoothest and faintest region of the image), then
 photometry is performed on each of the candidates, typically thousands per
 chip. The signal to noise for each detection is estimated by taking the ratio
 of the counts within a 0.5 pixel aperture to the rms noise in the sky background 
in the vicinity of the candidate. For this set of images we used a
 S/N cutoff of 3  to detect candidate clusters. We 
also used the ratio of the flux within an aperture of 0.5 and 3 pixels to weed
out chip defects and compact background galaxies. Using a concentration criteria
2 $<$ $\frac{Counts_{3pix}}{Counts_{0.5pix}}$ $<$ 13 for the PC and 2 $<$
 $\frac{Counts_{3pix}}{Counts_{0.5pix}}$ $<$ 10 for the WF, we identified  
cluster candidates that satisfied the selection criteria in both the F555W and
the F814W image. Some of the S0 galaxies of later type have ongoing star 
formation in the disk and our detection algorithm identified many spurious
 objects in or near the disks of these galaxies. As most of the candidates
in this region appear to be star forming regions, we
 masked out the point-like sources in the disks of NGC 1581, ESO 118-G034, 
NGC 2328 and NGC 3870.
  After correcting for the geometrical distortion (Holtzman {\it et al.} 1995a),
we performed aperture photometry  using a 3 pixel radius 
aperture for the PC and a 2 pixel radius  aperture for the WF, using the
median pixel value between 5 and 8 pixel radius as the sky background. 
 Since the profiles
of the cluster candidates are slightly broader than a stellar PSF, we used the
distance-dependent  aperture corrections computed in Paper I. To maintain consistency with Paper I  we adopted photometric
zeropoints of 22.573 mag and 21.709  for the PC and applied small
 color correction terms from 
Holtzman {\it et al.} (1995b) to convert the F555W and F814W magnitudes 
to Johnson V and  Cousins I respectively. Since the zeropoints for each of the WF 
chips are offset slightly from the PC values we added the small differences 
quoted in the {HST  data handbook (1997)} to the above values before applying the color correction 
terms to the candidate objects in the WF (see Paper I for details). We corrected for the foreground
Galactic extinction in the direction of each of the observed galaxies using
 the A$_B$ (Burstein \& Heiles 1982) values quoted in the NED extragalactic database and the reddening curve 
from Mathis (1990). The V-band extinctions are listed in Table 1. A careful observation of the V, I and V-I images revealed that internal reddening is not an issue in most of the galaxies in our sample. Significant dust lanes are only seen in the disks of the very late type S0s, but as indicated above we already exclude these regions because of confusion due to star forming regions. A few
of the earlier type galaxies have weak dust lanes in a very small region around the nucleus - of the order of a few arcsecs - which does not affect the global properties of the observed cluster systems.
We did
 not correct for the Charge Transfer  Efficiency (CTE) gradient across the chips
  as the CTE problem is expected to be minimal in the presence of the 
strong galaxy background in most of our sample  objects. (Holtzman {\it et al.} 
1995b; Whitmore \& Heyer 1997). Even in regions
of low background the CTE corrections are expected to be small ($\sim$0.02
 mag) at the epoch of these observations (Whitmore et al. 1999).

	For statistical studies of  globular cluster systems it is necessary to 
quantify the photometric incompleteness of our detections i.e. the ability to
 detect candidates as a function of magnitude. 
Since the exposure time for all of our images is identical, the completeness curve 
for each galaxy is expected to be only a function of the galaxy background.
This makes our task a little simpler in that we only need to measure the
detection threshold for a representative sample of galaxies. We selected a 
random set of galaxies that spanned a large range of background counts and 
added simulated clusters, which we then attempted to detect with the 
routine  used to identify the cluster candidates. We only added 100
 objects at random positions in each chip during each simulation - in both the 
F555W and F814W images -  to ensure that there was negligible overlap of objects. 
Moreover, since a preliminary examination of
 our data showed that the colors of most of the candidate globular clusters are
 in a narrow range of values around V-I$\approx$1.0 as expected for old globular clusters, we
allowed our simulated objects to have random colors between 0.8$<$V-I$<$1.3. In all, we added 
approximately 35000 objects using the IRAF task ADDSTAR. The completeness
curves derived from these tests are presented in Fig 1. The typical 50\% detection
threshold is V$\approx$23.4 mag in the PC, and  V$\approx$24 mag in the WF. The
I-band detection limits are $\approx$1 mag brighter than those in the V-band.
These values are comparable to the deepest ground-based observations, even though this analysis is based on $'$snapshot$'$ 160s exposures in the V-band and  320s in the I-band.

	Globular cluster systems of galaxies outside the local group can be
measured statistically as an overdensity of point-like objects as compared to a
 blank field a small distance from the galaxy. 
We do not have off-galaxy images for this set of observations, but the superior
angular resolution of the HST makes this almost redundant  since globular
 clusters can generally be distinguished both from unresolved foreground stars
and background galaxies.  A judicious color cutoff is used to further 
eliminate most of the contaminants. At low galactic latitudes, however, there
may be large number of  foreground stars in the field of view. Also, cluster systems with less than $\sim$50 clusters are likely to have significant contamination. We have noted the
 potential contamination problem in specific galaxies in the next section.

\section{Results and Discussion}

	The color-magnitude diagrams of the point-like objects in the program
galaxies are shown in Fig 2. A  majority of  the cluster candidates 
  lie in a narrow range of color  between 0.5$<$V-I$<$1.5 with 
a mean color near V-I $\approx$ 1.0 mag, which is typical for old GCSs.  

 In the galaxies which have significant cluster 
populations one can discern that the mean of the color distribution is roughly
constant with magnitude. The dispersion of the colors does seem to increase for
 the fainter objects, but this is most likely a reflection of the larger
 photometric uncertainties at these magnitudes. Galaxies at low galactic 
latitudes (NGC 404, NGC 1201, NGC 2328, NGC 2902, NGC 3056 \& NGC 6703) appear 
to have a larger fraction of objects redder than V-I = 1.5 mag, which can be
attributed  to contamination by foreground stars. In order to help filter 
out the few remaining contaminants, largely foreground stars and background galaxies, we
 shall consider only objects within the color range 0.5$<$V-I$<$1.5 to be bona fide
cluster candidates and all objects outside this range to be $'$background$'$
objects (meaning both foreground and background interlopers).

	We inspected the spatial distribution of the cluster candidates in each 
of the 34 galaxies and found that the density distribution of the cluster 
candidates in most of the galaxies was roughly centered on the nucleus of the 
galaxy, further suggesting
 that the candidates are bona fide clusters associated with the galaxy 
 in question. The cluster candidates of four galaxies (ESO 118-G034, NGC 2328,
VCC 165 and IC 3131) did not appear to be spatially correlated to the underlying
galaxy, which leads us to believe that either these galaxies do not have
 significant cluster systems within our field of view or they are heavily 
contaminated by  $'$background$'$ objects. We shall not discuss the globular 
cluster systems of these galaxies in this paper except in $\S$3.3 where we place
 limits on the specific frequency. NGC 404 presents a unique problem since the 
luminosity and color distribution look 
very different from those of the  other S0s, with a sudden increase in both
 the number of objects and the color spread of the candidates near V$\approx$24 
mag (or I$\approx$23.5).  We suspect that we are actually observing the tip 
of the red giant branch in this nearby galaxy.  If we adopt the Lee, Freedman 
\& Madore (1993) value of M$_I$ $\approx$ -4.0 mag for the absolute magnitude of
 the tip of the red giant branch the distance modulus to the galaxy is 
m-M $\approx$ 27.5 mag. Thus we suspect that NGC 404 is probably at a distance 
of $\approx$ 3 Mpc,
and not 10 Mpc as suggested by Wiklind \& Henkel (1990) from their arguments
based on the morphological characteristics and  molecular gas content of the
galaxy.

 Twelve of the 29
 galaxies  have more than 50 candidate clusters in the field of view of our images.  The other 17 may be severely affected by
small number statistics and/or contamination. In subsequent sections we 
have analyzed the cluster properties of all 29 galaxies with appropriate error
estimates, allowing the reader to make the subjective decision of which
galaxies have 'believable' cluster systems. We note that we have included some of the numbers derived for the NGC 4550 globular cluster system in Paper I in relevant sections of this analysis of S0 galaxies. Although NGC 4550 is a SB0$^0$, it was not observed as part of the snapshot survey. Unlike the snapshot data set studied in this paper,  the reduction and analysis techniques employed for the deeper NGC 4550 images were identical to the elliptical galaxies in Paper I. Hence we decided that it would be more appropriate to include it in Paper I. When we weigh the properties of S0s vis-a-vis ellipticals we take care to include NGC 4550 in the S0 sample.

\subsection{The Color  and  Metallicity Distributions }
	
	The V-I color distributions of the cluster candidates in the range 
0.5$<$V-I$<$1.5 are shown in Fig 3. The spread in GC colors in each
 galaxy is  V-I$\sim$0.5 mag. This dispersion in color cannot be attributed to
 the photometric errors alone (typically $\approx$0.15 mag), hence it represents an actual variation in broad-band color from cluster to cluster.

	The mean colors of the globular cluster candidates in the color range 
 0.5$<$V-I$<$1.5  are listed in Table 2. The error estimates quoted are the formal uncertainties in the means (Note - Throughout this
 paper we have used the following convention in quoting uncertainties: The number following the $\pm$ sign is the standard deviation, while a number in parentheses refers to the uncertainty in the mean).	Stellar population modeling shows that the color difference between 
globular clusters is dominated by age effects for clusters younger than $\sim$1
 Gyr old, and by metallicity effects otherwise (Worthey 1994;
 Bruzual \& Charlot 1993 etc). Since the GC color and luminosity (\S3.2) distributions of our sample are 
fairly typical of old ($>$$\sim$8 Gyr) cluster systems, we can safely attribute 
 the color spread predominantly to metallicity variations. The mean metallicities of the GCSs, adopting the transformation equation based on Galactic clusters from Kundu \& Whitmore (1998), are also listed in Table 2. However, we note that the faintest clusters (in the V-band) in 
some of the galaxies in Fig 2 are selectively redder. A corresponding color-magnitude 
diagram plot in the V-I, I plane would show that the faintest
clusters in the I-band are selectively {\it bluer}. This is an artifact of the 
cluster identification method which required that the cluster candidate be present in both the 160s V
 image and the deeper 320s I image. Since this bias only occurs at the faint 
end of the luminosity function where the completeness is low, the overall
effect on the mean color is expected to be quite small. In order to quantify
this uncertainty we calculated the mean colors for various cutoff magnitudes in
V and I. The average offset in the mean color of the raw data and the magnitude 
limited samples is only $\sim$0.02 mag.  Therefore  we are confident that the mean colors
 are largely unaffected by selection biases.

In Table 2 we also indicate the number of cluster candidates within the color 
range of typical clusters, N$_{cand}$, and the number of sources with colors between 0.0$<$V-I$<$0.5 or 1.5$<$V-I$<$2.0, N$_{bg}$. The point-like objects in the latter subset are likely to be largely contaminants. One might reasonably infer
that galaxies with large values of N$_{bg}$ also have significant number of contaminants in the color range of clusters. Therefore, in calculating the mean
color of GCs in S0s we only consider systems with N$_{bg}$/N$_{cand}$$<$0.25.
 The average color of our S0 sample is V-I=1.00$\pm$0.07 (0.01) mag which corresponds to a mean metallicity of [Fe/H]=-1.1$\pm$0.3 (0.07) dex.  This is very similar to  the value of V-I=1.03$\pm$0.04 (0.1) derived for ellipticals in Paper I.

	Up to this point we have been discussing only the average metallicities
of the cluster systems; the shapes of the color histograms may also hold vital 
clues about the formation history of S0 galaxies. One of the most significant 
 discoveries in globular cluster research in recent times
is that the broad-band color distribution of 25-50$\%$ of the early type galaxies
 studied are bimodal (e.g. we showed in Paper I that at least 50\% of ellipticals have bimodal GCS color distributions). This effect, which is 
usually interpreted as a bimodality in the metal abundance of the cluster systems, 
suggests that globular clusters have formed during two distinct epochs in the
metal enrichment histories of these galaxies. However there is much disagreement
about the physical mechanism that triggers the second burst of cluster formation.
   Schweizer (1987) and Ashman $\&$
 Zepf (1992) proposed that GCs may be formed during the interaction or merger
of galaxies. While Zepf $\&$ Ashman (1993) suggest that the bimodality in the color
distribution is a natural consequence of the merger scenario, Forbes, Brodie \&
Grillmair (1997) favor multiple phases of cluster formation during the 
collapse of a gas cloud into a galaxy. 

The search for bimodality in the  color distribution of globular cluster systems
  has  largely been limited to elliptical galaxies; little is known about the 
distributions in S0 galaxies and the implications on formation models. To date, 
 NGC 1380 \& NGC 3115 are the only S0 galaxies confirmed to have a bimodal color
 distribution  (Kissler-Patig et al.  1997; Kundu \& Whitmore 1998). Since 
these are also the only S0 galaxies in which a deep study of the color distribution of the GCS has been made it is unclear what percentage of S0 galaxies are bimodal.

	An inspection of the histograms of the color distributions in Fig 3 
reveals no candidates with strikingly obvious bimodality.  In order to objectively test whether the unbinned
  distributions have multiple modes we applied the KMM mixture modeling algorithm
 of Ashman, Bird \& Zepf (1994), which tests the suitability of modelling a distribution with multiple Gaussian sub-populations. Since 
the KMM algorithm is very sensitive to outliers we did not attempt to use it 
on systems with fewer than 50 globular cluster candidates, where small number 
statistics and contamination by background and foreground objects may lead to
 false confirmations of multiple peaks. KMM tests on the more cluster-rich
 systems revealed that only NGC 2768 has a statistically significant probability
of having a bimodal color distribution, with peaks at V-I=0.92 mag and 
V-I=1.12 mag. On the face of it, it would appear that the fraction 
of S0 galaxies with multimodal color distributions is much smaller than that of
elliptical galaxies, but we hasten to add that we are limited by the small
 number statistics and the photometric uncertainties arising from
 the short exposure times of this snapshot survey. Ashman et al.  (1994) show that for sample sizes of 100-200 objects the photometric 
uncertainties should be a factor of $\sim$2.5-3.0 less than the expected separation
between the peaks for the KMM algorithm to reliably detect or reject bimodality.
Given that  in most galaxies with  bimodal color distributions studied to
date the difference in the peaks is of the order of V-I$\approx$0.2 mag and 
that the typical mean uncertainty in color in our data sets is V-I$\approx$0.15
mag we do not expect to detect bimodality in many individual systems.

Based on a visual inspection of the color-magnitude diagrams of the clusters and the
KMM test results  the following galaxies appear to be promising candidates
for bimodality : NGC 1201, NGC 1332, NGC 3489, NGC 4459, and possibly NGC 6861. Although KMM tests suggest that two Gaussians fit the color distributions of
 these galaxies better than a single Gaussian at a 95\% level of confidence in these systems the homoscedastic (populations forced to have equal 
variances) and heteroscedastic partitions (populations allowed different variances) yielded red and blue population with significantly different partitions and mean colors (different by $\sim$0.05 mag). Given the low number statistics and the photometric uncertainties we do not consider these partitions to be very reliable.
In order to improve the number statistics and search for stronger evidence of bimodality we co-added the likely bimodal systems. In the top left panel of Fig 4 we plot the  co-added color distributions of NGC 1201, NGC 1332, NGC 3489, NGC 4459 and NGC 2768. Though we see no obvious sign of bimodality in this plot, it does not necessarily mean that the individual systems are unimodal. It is possible that the bimodal peaks are at different locations in
each galaxy. In order to somewhat alleviate this problem we shifted
 the color distribution of each of the individual GCSs to a common mean of
V-I=0 and then co-added them. The V-I histogram (Top right of Fig 4) still shows no convincing evidence of bimodality. A KMM test on the co-added,
 normalized distribution suggests that two Gaussians with (normalized) peaks at
V-I = -0.08 and 0.09, with roughly equal number of candidates in each Gaussian, fits the data better than a single Gaussian at better than a 95\% confidence level. However, a careful visual inspection of the distribution reveals no obvious peaks at these
 locations. Thus one may only conservatively conclude from the KMM tests that
 the color distribution is broader i.e. more flat-topped than a Gaussian. The 
fact that the standard deviation in the mean (normalized) color, $\delta$$<$V-I$>$ = 0.18 is larger than the mean of the photometric uncertainty of individual clusters, $<$$\delta$(V-I)$>$ further supports this conclusion. In order to further test the significance of the color spread in S0s we compare them with confirmed bimodal  elliptical GCSs. In the lower left panel of Fig 4 we plot the co-added, normalized color distribution of 3 Virgo galaxies, M87 (Kundu et al. 1999), 
NGC 4472 and NGC 4649 (Paper I). KMM tests indicate that the distribution is
 well described by Gaussians with peaks at V-I = -0.12 and 0.10 at a confidence level of 95\%. Next we selected the subset of the Virgo clusters with the exact same photometric uncertainty distribution as the co-added S0s  by matching each S0 cluster with a Virgo GC with the same - or as close as possible - computed photometric uncertainty (taking care that each Virgo globular cluster is selected only once). The mean difference in color uncertainties between individual S0 clusters and Virgo clusters is -4.1$\times$10$^{-5}$ $\pm$ 0.001,
confirming that the uncertainty distributions are indeed almost identical. The
color distribution of this Virgo subset is shown in the bottom right panel of Fig 4. Clearly the raw histogram does not show very strong evidence for bimodality. However, KMM tests confirm that the Virgo clusters with S0 error distribution are inconsistent with a unimodal distribution, and can be better 
fit by Gaussians with peaks at -0.11 and 0.11. This gives us more confidence in the KMM tests for the normalized S0 candidates and suggests that these S0s may
indeed have non-unimodal color distributions. However, we do note that although
 by definition the Virgo subset has the same $<$$\delta$(V-I)$>$ (0.13) as the
 S0 distribution, the standard deviation in the mean color of the Virgo globular
clusters is slightly larger (0.20 as opposed to 0.18). We conjecture that this indicates that even though the S0 systems may be bimodal the ratio of red to blue clusters is significantly different from the 1:1 mix typical of giant ellipticals.

Thus, considering the galaxies in our sample plus NGC 3115, NGC 1380, and NGC 4550 the only
 other S0 galaxies which good data on the color distribution exists, we conclude that at
 least 10$\%$ of S0 galaxies have confirmed bimodal color distributions, at the present lavel of photometric accuracy. In light of the discussion in the previous paragraph we are unable to definitively rule out bimodality in most of 
the other cases due to small number statistics and relatively large photometric errors. We suggest that at least another 10$\%$ of S0s may be bimodal.

\subsection{The Globular Cluster Luminosity Function }

	The shape and turnover magnitude of the globular cluster luminosity function has been found to be nearly constant over a wide range 
of galaxies and environments (Harris 1991). While the theoretical basis for this
phenomenon remains unclear, this is a remarkable result which implies that for any 
reasonable range of M/L ratio the underlying mass distribution of globular clusters is similar in most galaxies. This can be exploited to determine distances to  galaxies (e.g. Jacoby et al.  1992; Whitmore 1996).
 In Paper I we improved upon previous estimates of the GCLF parameters - M$_V^0$ = -7.41 (0.03), M$_I^0$ = -8.46 (0.03) and $\sigma$ $\approx$ 1.3 for a Gaussian fit - and showed that the GCLF is an excellent distance indicator for elliptical galaxies. There are lingering doubts that the turnover magnitude varies with Hubble type (e.g. Secker 1992), although Ashman, Conti \& Zepf (1995) (hereafter ACZ) ascribe these differences to metallicity effects.

	Most of the galaxies in our sample do not have enough clusters for us to
 determine the peak of the GCLF accurately so we choose to study only the 14
richest cluster systems. The luminosity functions of these galaxies in the V and
 I-band, sorted by the distance modulus in Table 1, are plotted in Fig 5. In each case we have plotted the GCLF only for the candidates
that are fainter than V=(m-M)-11.0 in order to  limit the effect of outliers
- most likely foreground stars - on the luminosity function. 

	It is apparent from Fig 1 the completeness curves are background dependent
 and also vary between the WF and PC chips. Following the prescription laid out in detail in Paper I we computed a background count and cluster density weighted
 completeness curve for each of the candidates. The 50\% completeness limit in most galaxies is at V$\approx$24 and I$\approx$23. The completeness corrected GCLFs, up to the
respective 50\% completeness limit of each GCS, is shown by the dotted lines in Fig 5. As in Paper I we calculated the parameters of the best fit Gaussians to
the GCLFs by using the IRAF task 
NGAUSSFIT and varying the width (0.18 to 0.25 mag) and positioning of the bins,
 the cutoff magnitude at the faint end (40$\%$ to 55$\%$ completeness), and 
$\sigma$ (1.1 to 1.5). For the adopted mean $\sigma$ of 1.3 we calculated the
average turnover magnitude, the amplitude of the Gaussian and the associated 
standard deviation from 40 individual fits. The best fit Gaussian ($\sigma$=1.3)
GCLF is plotted using dashed lines in Fig 5 and the turnover magnitude is listed in Table 3. The uncertainties in the turnover reported in Table 3 are the standard deviations from the mean of the 40 fits to the GCLF and are a  good measure of both the systematic and formal uncertainties.

By and large, the peak of the cluster distribution migrates
to fainter magnitudes for more distant galaxies in Fig 5. We point
out that the distances quoted are from Table 1, which have have been derived from the velocity and are not likely to be very accurate.  In Fig 6 we plot the turnover luminosity in the I-band vs that in the 
V-band. It is apparent that m$_V^0$ and m$_I^0$ are tightly correlated, and within the uncertainties follow a linear relationship. The uncertainties are larger for the more distant, fainter systems as is only to be expected from
the completeness limits.

	In order to test the suitability of S0 GCLFs as distance indicators, we would ideally like to compare the turnover magnitude with distance moduli obtained by other methods and measure the constancy of the turnover. Unfortunately only a couple of galaxies in our sample have reliable independent  measurements using other distance indicators (see below).   Therefore, we opt instead, to adopt the absolute turnover magnitudes derived for ellipticals in Paper I and check for consistency between the derived S0 GCLF distance moduli and other distance estimates. The distance moduli to some of the galaxies reported in the literature are listed in
column 5 of Table 3. Most of the reported distances are from Prugniel \& Simien  (1996). As in Table 1, these distances are based on recessional velocities but they incorporate Virgocentric infall and great attractor corrections using an
independent model. For NGC 1400 we report the average fundamental plane distance modulus in the I (31.67$\pm$0.14) and K'(32.07$\pm$0.25) band from 
Jensen, Tonry \& Luppino (1998). Columns 6, 7 and 8 of Table 3 show the computed GCLF distances in the V, I and mean distance respectively, for the adopted Paper I turnover magnitudes of M$_V^0$=-7.41$\pm$0.03 and M$_I^0$=-8.46$\pm$0.03. We do not compute distances for
galaxies with mean GCLF uncertainties greater than 0.3 mag as these are likely to be very unreliable. The GCLF distances appear to be consistent with other distance estimates for our sample of S0s, and by extension there seems to be no obvious offset between the absolute turnover magnitudes of ellipticals and S0s. The mean difference between the GCLF distance moduli and those from the literature (column 5 of Table 3) is 0.01$\pm$0.4, quite similar to the difference between the distances from Table 1 and those reported in the literature, 0.07$\pm$0.5.
More accurate independent distance estimates of S0 galaxies and deeper observations of GCLFs are needed to further establish the accuracy (or applicability) of the turnover magnitude as a distance indicator in S0s. However, we note that the turnover magnitude of NGC 3115, the S0 galaxy with the best determined GCLF to date, yields a distance measurement that is compatible with other estimates and falls within the range of values derived by the tip of the red giant branch, planetary nebula luminosity function and surface brightness fluctuation methods (Kundu \& Whitmore 1998).  

	Ashman, Conti \& Zepf (1995 hereafter ACZ) showed that for a universal globular cluster mass function the peak of the GCLF is slightly dependent on the
metallicity. Both the V and I-band turnovers migrate to fainter magnitudes in more
metal-rich systems, the effect being larger in V. In Paper I we showed that 
 m$_V^0$-m$_I^0$ does increase with the color (metallicity) of the cluster system, consistent with the ACZ estimates. Fig 7 is a corresponding plot of m$_V^0$-m$_I^0$ as a function of mean cluster color for the S0 systems for which
GCLF distances are reported in Table 3. For comparison we also plot a point for
M87 and the best fit line from our elliptical sample, (m$_V^0$ - m$_I^0$) = 0.11 + 0.90$\times$(V-I). Rather surprisingly,
despite the large uncertainties in the turnover magnitudes derived from the 
shallow images of S0s, within the uncertainties the S0 sample is consistent with the  metallicity trend of the elliptical sample. This indicates that systematic errors
dominate the GCLF determinations of the short exposure S0s, and that these
errors are correlated in the V and I-band, i.e. Given that the ratio of exposure times in F555W \& F814W is exactly the same for the entire S0 data set, the ratio of the fraction of the V-GCLF vs the I-GCLF used for our Gaussian fits is roughly the same for galaxies at all distances in our sample. We suggest
that the color dependence of m$_V^0$-m$_I^0$ indicates that the shape of the bright end of the GCLF is similar in both the V and I-bands. Therefore even in the most distant candidates in which the 50\% completeness limit falls far short of the turnover magnitude the difference in m$_V^0$-m$_I^0$ can be determined 
more accurately than the individual turnovers in either filter. This would also explain why m$_V^0$ and  m$_I^0$ appear extremely well correlated, despite the large uncertainties, even for the faintest GCSs.

	Based on the analysis of the globular cluster luminosity functions of our S0 sample we conclude that within the uncertainties, the turnover magnitude of S0s are consistent with other distance measurements, and that the bright end of the GCLF luminosity function has a similar shape in the V and I-band.

 \subsection{Specific Frequency }

	The specific frequency, S$_N$, is defined as the number of globular 
clusters per unit galaxy luminosity normalized to M$_V$=-15 (Harris \& van den Bergh 1981):

 S$_N$=N$_t$ 10$^{-0.4 (M^T_V + 15)}$
   
where N$_t$ is the total number of clusters and M$^T_V$ is the total magnitude 
of the underlying galaxy. Given the limited field-of-view of the WFPC2 
our images only cover a portion of the sample galaxies. Therefore we cannot 
measure the global specific frequency of the galaxy directly. We can, however, derive the
local value of S$_N$ within the HST field-of-view  using
 the total V-band magnitude of the galaxy in our image and
the projected total number of clusters in the region. Our method of calculating 
these two quantities and the underlying assumptions are outlined below.

	The total number of candidates with colors in the range 0.5$<$V-I$<$1.5 
 detected in each galaxy and the number of
objects with colors in the range 0$<$V-I$<$0.5 or 1.5$<$V-I$<$2.0 are listed in Table 2 
under the headings N$_{cand}$ and N$_{bg}$ respectively. We then calculated 
histograms of the candidate objects brighter than V=24 mag and corrected them by the 
 completeness curves derived in the previous section to arrive at the projected total 
number of candidate 
objects with V$<$24 mag within our field-of-view, N$_{V<24}$. This  number  is roughly the same as N$_{cand}$ since the number of 
objects fainter than V=24 mag is small and this is effectively offset by the completeness correction. 

Next we must account for the contamination by background and foreground sources.
 Since we do not have off-galaxy images we attempt to correct the contamination in 
color space. For the 4 galaxies which appear to have no cluster candidates
we found that  N$_{V<24}$ is roughly equal to 1.5$\times$N$_{bg}$. Assuming 
that this is a typical background in all our galaxies we subtract 1.5$\times$N$_{bg}$ 
from N$_{V<24}$ to arrive at the total number of 
contamination and completeness corrected objects brighter than
V=24 mag in our field-of-view. As in the previous section, we adopt the parameters of a Gaussian GCLF describing elliptical galaxy GCSs, $\sigma$=1.3 and a turnover luminosity of M$_V^0$=-7.41. Using the 
distance estimates from Table 1 to calculate the expected fraction of clusters
brighter than M$_V$=24 mag and thence the projected total number of clusters 
within our field-of-view, N$_{Tot}$. We estimate that the uncertainty in the 
total number of clusters is the sum in quadrature of the uncertainty from Poisson statistics, $\sqrt{N_{cand}}$, and the uncertainty due to contamination,  N$_{Tot}$ $\times$ ($\frac{N_{bg}}{N_{cand}}$). As a check, we calculated the integrated area under the GCLF and the associated errors for the candidates
with well defined luminosity functions in the previous section, in a manner similar to the Paper I analysis. While the total number of clusters returned by both methods was similar, the uncertainty in the total number of clusters was 
$\sim$2 times as large using the latter method. Thus the uncertainties in the total number of clusters, and hence the uncertainty in the local value of S$_N$ is {\it typically underestimated by a factor of 2} in this paper when compared to the analysis of the elliptical sample.

	Determining the integrated magnitude of the galaxy is more
 problematic because  nearly all of our galaxies are spatially larger than the
 WFPC2 field-of-view, making background subtraction impossible. Moreover, we
have only one F555W (V-band) image for each galaxy which means that the cosmic 
ray rejection is far from ideal. While this does not affect aperture photometry of
the point-like globular clusters - the likelihood of a cosmic ray event being
superposed on a point source is very slim -  surface photometry can be
 significantly degraded. Instead, we calculated the integrated light within the
F814W image, subtracted 'the typical sky background' in the I-band  from the {\it WFPC2 Handbook} (0.0544 e$^-$s$^{-1}$pixel$^{-1}$ in the WF and 0.011 e$^-$s$^{-1}$pixel$^{-1}$ in the PC - WFPC2 Handbook)
 and adopted a value of V-I=1.2$\pm$0.1, which is fairly typical of 
S0 galaxies, to determine the integrated V magnitude of the program galaxies. We assumed that the uncertainty in the surface photometry is of the order of the subtracted $'$background$'$, plus an additional 0.1 mag due to the transformation from I to V and propagated this in the calculation of the uncertainty in the specific frequency. In order to check the accuracy of the surface photometry we compared our derived V-band photometry in small apertures with published aperture photometry. For an aperture of 29.8$''$ in NGC 6861 we derived V = 11.84 mag, compared to the published value of 11.88$\pm$0.02 mag (Sandage \& Visvanathan 1978). For apertures of 25.9$''$, 29.8$''$ and 35.6$''$ in NGC 1553 we calculated V = 11.16 mag, 11.04 mag and 10.91 mag, in good agreement with the published values of 11.19$\pm$0.02 mag, 10.94$\pm$0.02 mag and 10.84$\pm$0.03 mag respectively (Persson, Frogel \& Aaronson 1979; Sandage \& Visvanathan 1978; Sandage 1975). The published numbers of V = 11.53$\pm$0.02 mag and 11.46$\pm$0.02 mag (Persson et al 1979; Sandage \& Visvanathan 1978) for 25.9$''$ and 29.8$''$ apertures are consistent with our values of 11.45 mag and 11.36 mag in NGC 1332. Thus the uncertainties in the integrated magnitude of the hosts adopted by us seems to be reasonable estimates.

	The local specific frequency in the inner region of the galaxies is
 listed in Table 2 along with the global values for the candidates that have 
previously been observed in ground-based studies. The local specific frequency varies
 from 0 for galaxies with no measurable cluster systems to 6.8 for the
 cluster rich dwarf NGC 3115 DW1, with a mean value of 1.0$\pm$0.6 (0.1) for all S0s with uncertainty in individual local specific frequencies, $\delta$S$_{N(Local)}$, less than 1.5. This is lower than the local specific frequency of 2.4$\pm$1.8 (0.4) derived for elliptical galaxies in Paper I and is consistent with previous observations that the specific frequency of S0s is lower than that of ellipticals (Harris 1991).  The large spread in S$_N$ from galaxy to
galaxy is real, and any unified theory of globular cluster formation must be 
able to account for vastly different efficiencies of globular cluster formation
in individual galaxies. In this connection NGC 3115 DW1 is an especially interesting case. As has previously been pointed out (Hanes \& Harris 1986; Durrell et al. 1996; Puzia et al. 2000) this dwarf galaxy appears surprisingly cluster rich. This can also be
clearly observed in the  color-magnitude plots in Fig 2 which are sorted by the 
absolute magnitude of the galaxy. While the density of clusters roughly
 increases with the brightness of the host galaxy, NGC 3115 DW1 has an
anomalously overdense system as compared to other galaxies of similar size. 

The local specific frequencies derived in both Paper I and this analysis are of course measured within the WFPC2 field-of-view, hence a larger fraction of the cluster population and galaxy light is included for
the more distant galaxies. While specific comparisons between individual galaxies in either sample may not yield very meaningful results, on average both the S0
 and the elliptical hosts are distributed over a similar distance range [(m-M) = 31.4$\pm$0.7 for ellipticals and (m-M) = 31.2$\pm$1 for S0s], hence we can legitimately make comparisons between the average properties of both samples. 
Also, in consort with published values, it is possible to study the radial properties of the specific frequency of each galaxy. The local specific frequencies calculated here add to this reference list.

 4 of the 5 galaxies  which have ground based estimates of S$_N$ all have smaller local specific 
frequency estimates in the inner regions  than the global value.  Furthermore, Fig 8 shows that the ratio of
 S$_{N(Local)}$/S$_N$ increase as a function of the absolute magnitude of the host galaxy. This is consistent with  other analyses of ellipticals (Forbes
et al.  1996; Grillmair, Pritchet \& van den Bergh 1986; Kundu et al.  
1999) and S0s (Kundu \& Whitmore 1998) that suggest that the globular cluster density distribution near the center of a galaxy is flatter than the stellar light distribution.  Although NGC 3115 DW1 appears to be an exception in that the local specific frequency is larger than the global value, we point out that due to the small angular size of
 this dwarf the WFPC2 image likely subtends a large fraction ($\sim$80\%) of both the galaxy light and cluster system of this host. Thus, we are effectively
measuring the global specific frequency of NGC 3115 DW1. The difference between
the effective global S$_N$ calculated here and the smaller specific frequency reported in the literature may be due to larger statistical uncertainties in
background subtraction in the earlier ground-based analysis.
The roughly linear relationship between  S$_{N(Local)}$/S$_N$ and M$_V^T$ is largely due to the fact that a larger fraction of the cluster system of smaller, less massive galaxies is subtended within the WFPC2 field-of-view.
  The elliptical sample in Paper I shows a similar effect. Thus, based on our analysis of the specific frequencies in the inner regions of ellipticals and S0s it is apparent that S0s are significantly less cluster rich than ellipticals.

\section{Globular Cluster vs Host Galaxy Properties }
	Now that we have established the broad characteristics of the GCSs (Table 2) 
we turn to  questions about their
relationship with the parent galaxy and the implications on various formation
and evolutionary scenarios. In Fig 9 we plot the average metallicity of the 
cluster systems (adopting the V-I to Fe/H transformation equation from Kundu \& Whitmore 1998) vs the absolute magnitudes
 of the host galaxies. We have not plotted the values for GCSs with N$_{bg}$/N$_{cand}$ $>$ 0.25  as the color estimates for these are likely
 to be affected by contaminating sources.  The magnitudes plotted along
 the horizontal axis are {\it not} the 
values listed in Table 2 but the integrated magnitudes of the entire galaxy
obtained from the NED catalog, normalized by the Table 1 distances. We have also included WFPC2 data points for NGC 3115 (Kundu \& Whitmore 1998), the elliptical sample from Paper 1, and M87 (Kundu et al.  1999), all of which have V-I color measurements. At face value Fig 9 suggests that there is an overall tendency for brighter S0 galaxies to have more metal rich cluster systems.  

The existence of the magnitude-mean metallicity
relationship has been a rather contentious issue and there are conflicting 
reports in the literature. While van den Bergh (1975), Forbes et al.  (1996) and
Bridges et al.  (1997) claim that the GCS metallicity-galaxy luminosity 
relationship is valid for all types of galaxies, Ashman \& Bird (1993)  suggest 
that elliptical galaxies show no perceptible trend. However, our analysis of ellipticals in Paper I, and Fig 9, confirm the existence of such a trend in both ellipticals and S0s, with no obvious offset between the two. A linear fit 
{\it to only our S0 sample} yields the expression.

 [Fe/H] = -4.9($\pm$0.6) - 0.19($\pm$0.03) M$_V^T$

The stars in the host galaxies themselves show a similar relationship between Fe/H and M$_V^T$ (Brodie \& Huchra 1991).
 The implied fact that
the galaxy metallicity and the GC metallicity are correlated  is not surprising 
in itself as it simply suggests that cluster formation is accompanied by star
formation in the galaxy.  Though neither the merger model nor the multiple collapse model
explicitly predict such a relationship between the metallicity of the globular cluster distribution and the absolute magnitude (mass) of the host galaxy, it is not inconsistent with either model. The similar spread of S0 and elliptical GCS metallicities in Fig 9,  and the almost identical mean metallicities of our S0 and elliptical samples do however suggest that the metallicity of the cluster system of an individual galaxy is primarily a function of the magnitude (mass) of the host galaxy and not of its Hubble type.

	In  the  top panel of Fig 10 we plot the local specific frequency  vs the mean color of the cluster system for the S0 and elliptical samples (restricted to S0 systems with $\delta$S$_{N(Local)}$ $<$ 1.5 and elliptical systems with $\delta$S$_{N(Local)}$ $<$ 3 for the reasons outlined in the previous section). There is no obvious trend in the local specific frequency of 
S0s with mean cluster color. Even for elliptical galaxies, in which the global specific frequencies have been shown to be weakly correlated with color (Ashman \& Zepf 1998 and references therein), we see no obvious trend in the local specific frequencies. Thus, the local specific frequency which is sensitive to the field-of-view, and possibly the effects of destruction mechanisms in the inner regions of galaxies, is not very helpful in discerning the efficiency of cluster formation at various metallicities. Further studies of global
 specific frequencies are needed to establish whether metal-rich S0 systems are more cluster rich,  and the impact of the results on the various models of S0 formation e.g.  minor mergers (Schweizer 1990),
 multiple collapse  (Forbes et al. 1996), or stripped spirals (Spitzer \& Baade 1951; Gunn \& Gott 1972).

	In the bottom panel of Fig 10 the local specific frequency of S0s and ellipticals are plotted as a function of host magnitude. While there is no obvious trend in the elliptical or S0 samples individually, there is evidence that the local
specific frequency of ellipticals at all host galaxy absolute magnitudes (masses) is slightly higher than that of S0s. This effect is more apparent for the less massive galaxies. For the most massive hosts a much smaller fraction of the total cluster system - and indeed of the galaxy - is visible within the HST field-of-view, and the local specific frequency {\it may} be influenced more by 
destruction effects than by global host galaxy morphology. Furthermore, considering the selection effect of the brightest galaxies  in our sample(s)
 being  spread out over a larger range of distances, coupled with the effects of
 calculating S$_N$ within the fixed WFPC2 field-of-view, we do not consider the apparent lack of an offset in the local specific frequencies of the brightest ellipticals and S0s to be a very secure or significant result. 

	It is well known that specific frequency is a function of Hubble
 type; early type galaxies have richer globular cluster systems than 
later ones (Harris 1991). With our large sample of S0 galaxies we can attempt to
 check whether there are any trends within S0s. Preferably, one would 
like to define the position of a particular galaxy on the Hubble sequence 
using a quantitative estimator like the disk-to-bulge ratio. Unfortunately, the
 disk-to-bulge ratio has not been calculated for many of our galaxies, and it is
not possible for us to determine it based on our own data due to the limited
 field-of-view. We attempted to bin the galaxies by their morphological 
classification in Table 1 but we soon discovered that some of the classifications were at odds with the structures seen in our images. We finally decided to roughly
reclassify the objects ourselves after visually inspecting the images. We 
consider all galaxies which have star forming regions or obvious structure in 
the disk to be of S0$^+$ type, galaxies that have a structureless disk to be of S0$^0$ type; the remainder we classify as S0$^-$. Many of the S0$^-$ galaxies 
may be more accurately classified as disky ellipticals, nevertheless we use the 
classification scheme outlined above for the purpose of this study. Fig 11 plots
 the variation of S$_{N(Local)}$ as a function of galaxy type for systems with $\delta$S$_{N(Local)}$ $<$ 1.5. 
While there is a statistically insignificant difference between the average 
specific frequencies of S0$^-$ [1.03$\pm$0.7 (0.2)] and S0$^0$ [0.95$\pm$0.6 (0.2)] galaxies,  S0$^+$s [0.7$\pm$0.4 (0.3)] appear to be significantly less 
cluster rich than than the early types. Hence, our observations are in agreement with the general trend that later type galaxies are less cluster-rich
and suggest that the average specific frequency may vary smoothly across Hubble type. Finally, we note with interest  that a significant fraction of S0s of all types
 actually have a genuinely low specific frequency. While much effort has been 
spent in trying to understand why certain
galaxies have inordinately large values of S$_N$, the equally intriguing, but neglected problem of why
some galaxies genuinely lack clusters might hold as many important clues to understanding the formation and evolution of cluster systems.

\section{Globular Cluster System vs Galaxy Cluster Properties }

 	There are indications that the properties of globular cluster systems 
may be correlated with the galaxy cluster that the host resides in. Blakeslee (1997) 
found that the number of globular clusters around the central galaxy in 
his sample of Abell clusters is proportional to the velocity dispersion
(or mass) of the Abell cluster. There have also been suggestions that the 
specific frequency of clusters in a galaxy is linked to the position of the galaxy within
the host cluster, with more centrally located galaxies having richer cluster systems. 

	Our (sample of) galaxies reside in widely disparate environments. Some of 
them are field galaxies, some are in poorly defined groups and others in well
defined galaxy clusters. Although we cannot determine the position of the host
galaxy within a cluster for most of our candidates as they are in 
poorly defined groups, we can attempt to define the total mass of the 
galaxy cluster system and look for correlations with GC properties. The mass of a galaxy cluster is usually best 
determined from the X-ray temperature. Unfortunately, very few of our sample
galaxies reside in clusters that have X-ray temperatures (or any X-ray 
gas for that matter). Instead we use the number of bound galaxies in the 
host cluster of each program galaxy, as determined by the LGG classification 
of Garcia (1993) as an estimate of the mass of the host galaxy cluster. This
is obviously not a very satisfying way of estimating the mass of the galaxy 
cluster but it allows us to very roughly differentiate small galaxy clusters
from large one.

 The  local specific frequency and average color of the elliptical and S0
 systems are plotted as a function of the logarithm of the number of 
galaxies in the host (galaxy) cluster in Fig 12. Note that the algorithm used by
 Garcia (1993) is sensitive to subclustering within  large galaxy clusters like 
Virgo and Fornax, hence the number of galaxies in these clusters  appear to be smaller than one would expect.  There is no obvious trend in the 
specific frequency or color of the GCS with galaxy cluster properties.
 We also traced the positions of the galaxies in the Virgo and Fornax
clusters on maps of the respective galaxy clusters to see if there is any correlation with position. While the complicated structure of the Virgo cluster makes the determination of its boundaries quite difficult, the Fornax
cluster is much more compact and better defined. While NGC 1375 and NGC 1389
appear to be fairly close to the core of the Fornax cluster, IC 1919 and ESO 358-G034 lie on the outskirts of the cluster. The similar, low specific frequencies of all 4 galaxies reveal no correlation between the specific
 frequency and location within the cluster.

\section{Sizes of the Clusters }

  As mentioned in $\S$2 the profiles of the globular clusters in our galaxy sample
  are slightly broader than stellar profiles, indicating that they are spatially
 resolved. Using the technique described in Kundu \& Whitmore (1998) we measured the sizes of the clusters. The median half-light radius of the clusters in the PC,
for galaxies that have at least five candidates in the chip, are listed in Table
 4.  The distance estimates from
Table 1 have been used to convert the angular size to parsecs. 

	The half-light radii (r$_h$) of individual clusters in the PC  - using
 Table 1 distances -  are plotted against cluster magnitudes for
 each of the galaxies in Fig 13. As in the elliptical sample in Paper I we find that the half light radius of clusters fall within a fairly narrow range of 0 - 6 pc in all cases with no apparent variation with magnitude. The large scatter in the sizes of the clusters in NGC 6861 is due to the fact that the S/N is the lowest in this the most
 distant galaxy of our sample.  There is a small trend of increasing mean r$_h$
with distance to the galaxy which, for the reasons outlined in Paper I,  we suspect is due to small errors in the adopted PSF model. The mean half-light
radius of our S0 sample is 2.6$\pm$0.7 (0.26), which is consistent with the 2.4$\pm$0.4 pc mean size of clusters in elliptical galaxies derived in Paper I and the 3 pc typical half-light radius of globular clusters  in
the Milky Way (van den Bergh 1996). 

\section{Summary}
	We have studied the cluster systems of 34 S0 galaxies from short exposure, 'snapshot' WFPC2 images in the V (F555W) and I (F814W) bands. Of these 34 host galaxies, 29 have measurable globular cluster systems. In each galaxy we have detected a population of old  globular clusters with colors
 in the range 0.5$<$V-I$<$1.5 mag.  

 The mean color (metallicity) of the GCSs of individual galaxies 
varies from V-I=0.85 mag ([Fe/H]=-1.88 dex) to V-I=1.17 mag ([Fe/H]=-0.37 dex).
The average metallicity  of S0 GCSs is [Fe/H]=-1.1$\pm$0.3 dex (V-I=1.00$\pm$0.07 mag) which is very similar to that derived for ellipticals in Paper I [Fe/H]=-1.0$\pm$0.2 dex. The mean metallicity of the cluster system of both S0s and ellipticals appears to be  primarily a function of the absolute magnitude (mass) of the host galaxy with no apparent dependence on the Hubble type.  

On the other hand, the local specific frequency of S0s is 
smaller than that of ellipticals. The mean  S$_{N(Local)}$ for S0s, 1.0$\pm$0.6 (0.1) is significantly smaller than the value of S$_{N(Local)}$ = 2.4$\pm$1.8
 (0.4) derived for ellipticals in Paper I. The
local frequencies of very late type S0 galaxies (S0$^+$) is
statistically lower  than that of early types (S0$^-$ and S0$^0$) which further
suggests that the mean specific frequency is a function of Hubble type.

Although NGC 2768 is the only galaxy in our sample with a confirmed bimodal metallicity distribution statistical tests indicate that several other galaxies are promising candidates for bimodality e.g. NGC 1332 and NGC 1553. Including S0 galaxies previously studied, we estimate that at least 10$\%$-20$\%$ of S0 galaxies  have bimodal cluster distributions at the present level of photometric uncertainty. 

The average half light radius of globular clusters in S0 galaxies is 2.6$\pm$0.7 pc. which is similar to that of ellipticals (Paper I) and the Milky Way.

 There are no significant trends in the mean metallicities or specific 
frequency of GCSs with membership of the host  in a galaxy cluster. We also find no evidence of a dependence of globular cluster properties with the location of the host  within the galaxy cluster. 

For the richest few S0 systems for which we derived the turnover luminosity, the GCLF distances were in good agreement with other estimates. As in the elliptical sample we find evidence that the difference in the turnover luminosities in V and I increases with metallicity. In Paper I we found that the GCLF turnover magnitude of ellipticals is an excellent distance indicator with an accuracy comparable with the surface brightness fluctuation method ($\sim$0.1mag). There is every likelihood that deeper images of S0s, and more accurate comparative distance estimates using other indicators, will reveal a similar level of confidence in the use of S0 GCLFs as distance indicators.

	AK is grateful to Mike A' Hearn, Francois Schweizer, Sylvain Veilleux 
and Stuart Vogel for numerous suggestions that helped improve this paper in its 
earlier incarnation as a chapter of his thesis. Yan Fernandez for all his 
suggestions and help. The authors would also like to thank the  referee Doug Geisler for his useful comments and suggestions. Support for this work was provided by NASA through grant number AR-08378.01-97A from the Space Telescope Science Institute,
 which is operated by AURA, Inc., under NASA contract NAS5-26555. This research 
has made use of the NASA/IPAC Extragalactic Database (NED) which is operated by 
the Jet Propulsion Laboratory, California Institute of Technology, under
 contract with the National Aeronautics and Space Administration.

\begin{figure}
\centerline{\psfig{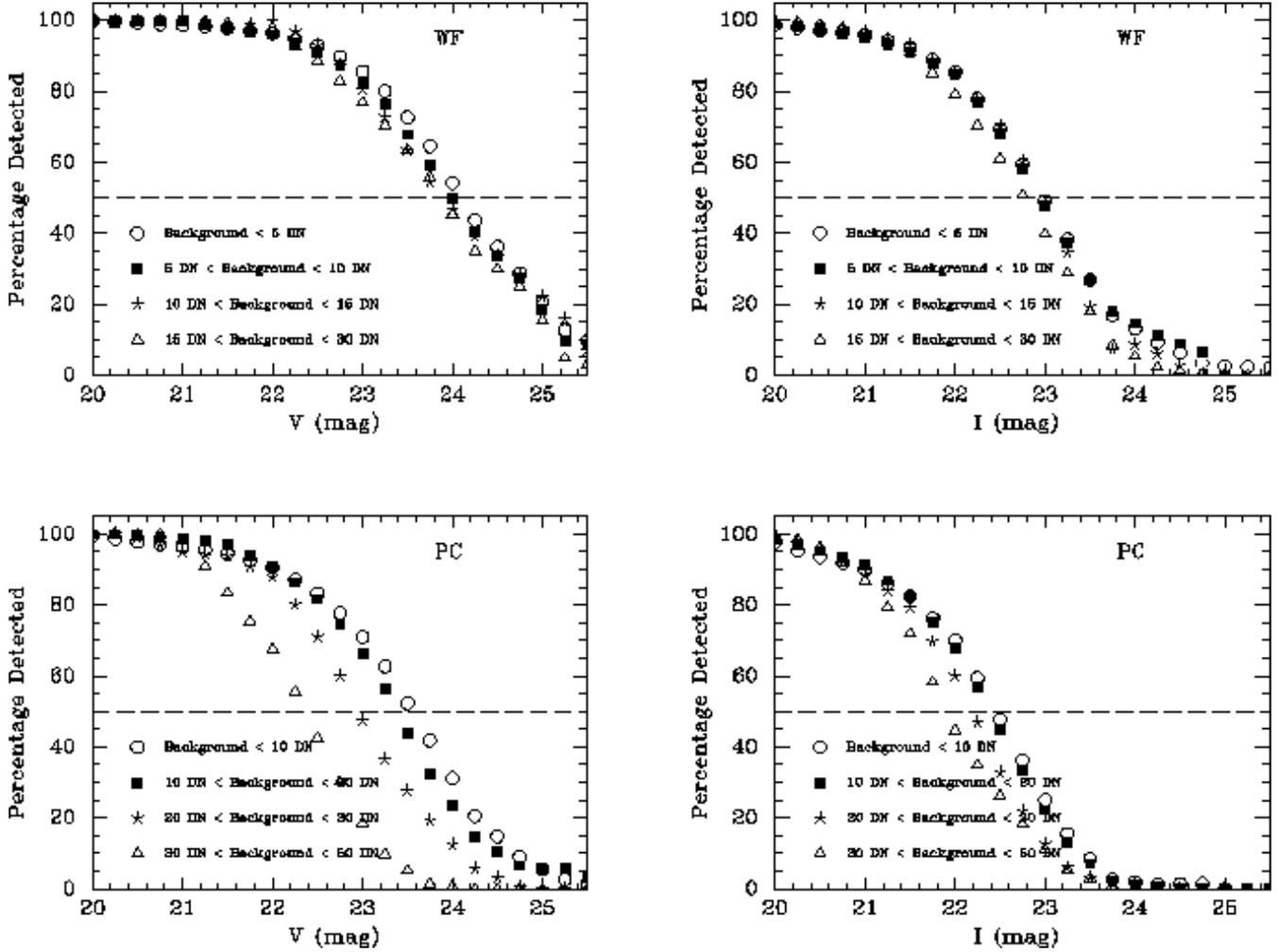}}
\caption{ Completeness curves for globular clusters in the V (F555W)  and I (F814W) images  for
objects with a V-I color of $\sim$1.0 mag as a 
function of background counts (see text for details). Even though the V-band 
images are only 160s snapshots, the 50\% completeness limit is at about V=24 mag, comparable to deep ground-based images. \label{fig1}}
\end{figure}

\begin{figure}
\centerline{\psfig{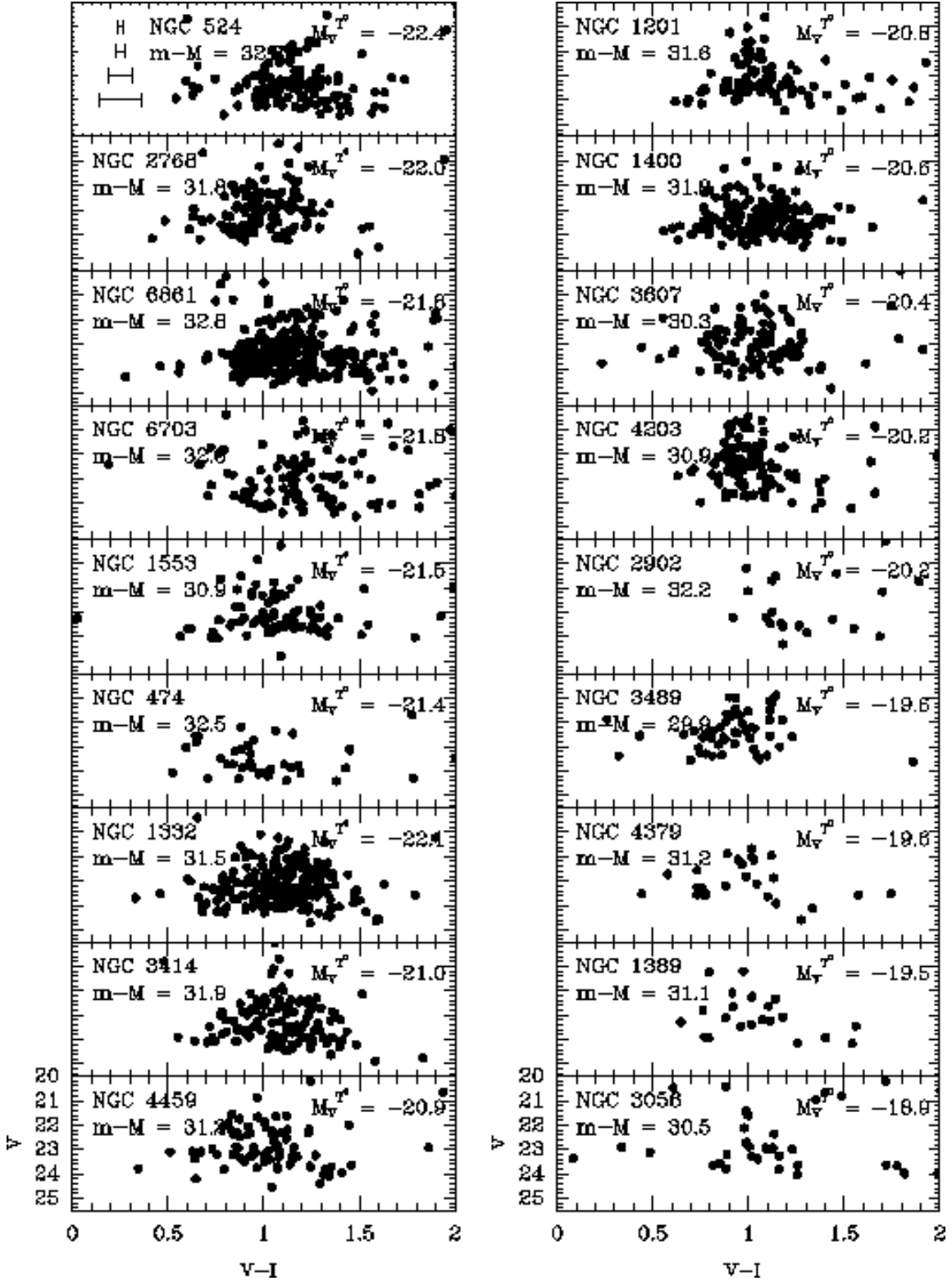}}
\caption{Color-magnitude diagrams for the globular cluster candidates in the
program galaxies. Most of the candidates lie in a narrow range of color between
0.5$<$V-I$<$1.5. The distances (from velocities) and  magnitudes are from Table 1. The galaxies are sorted by the absolute magnitude of the host. \label{fig2}}
\end{figure}
\clearpage
\centerline{\psfig{figure=fig2b.epsi,width=6in}}
\clearpage

\begin{figure}
\centerline{\psfig{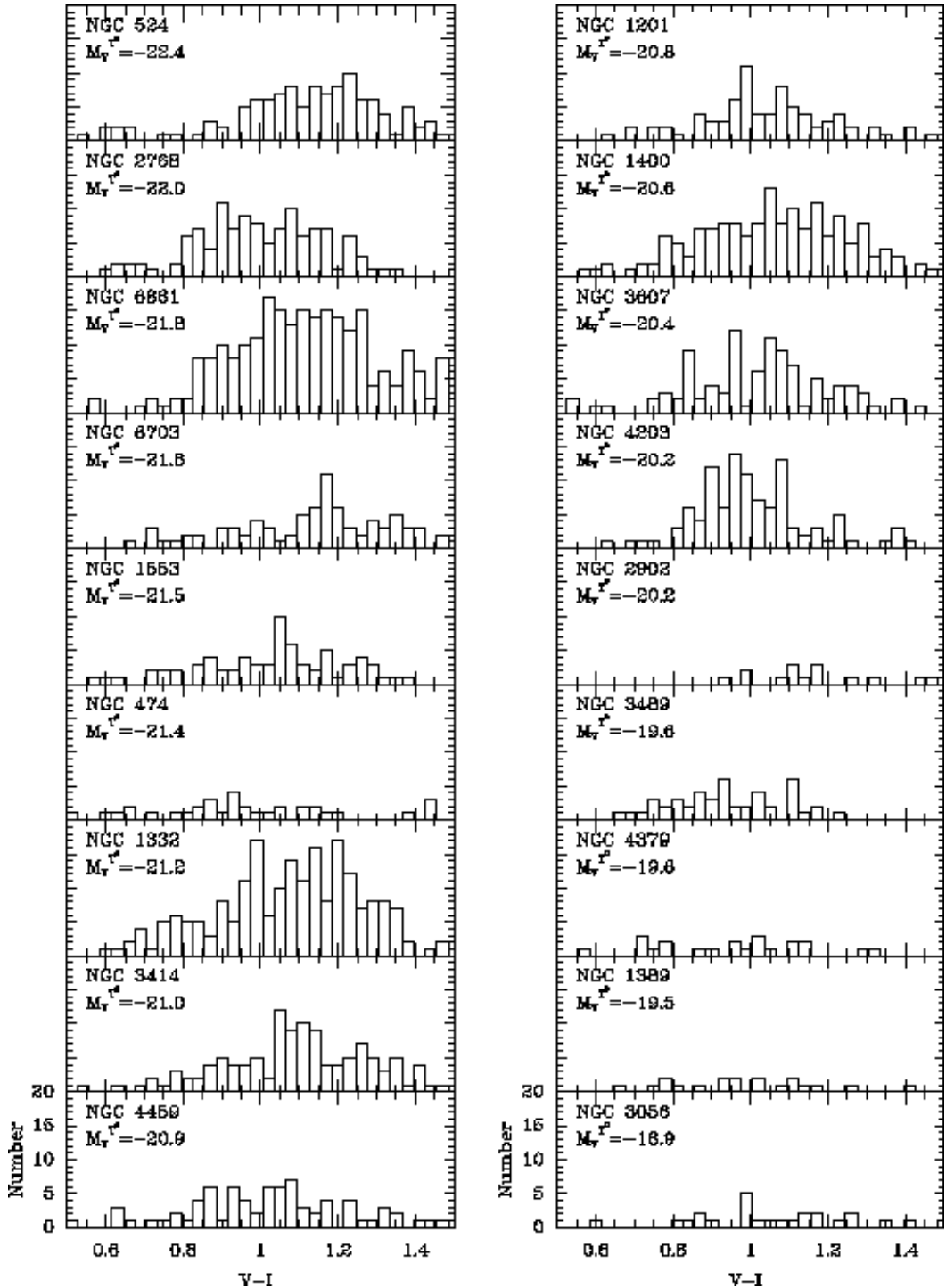}}
\caption{Color distributions of the globular cluster candidates in the 29
galaxies with measurable cluster systems. Statistical tests indicate that
NGC 2768 is the only galaxy in our sample with a confirmed bimodal color distribution.
 The small number of cluster candidates and the uncertainties in the photometry arising from the short exposure times  make it impossible to statistically confirm, or rule out, bimodality in most cases. \label{fig3}}
\end{figure}
\clearpage
\centerline{\psfig{figure=fig3b.epsi,width=6in}}
\clearpage

\begin{figure}
\centerline{\psfig{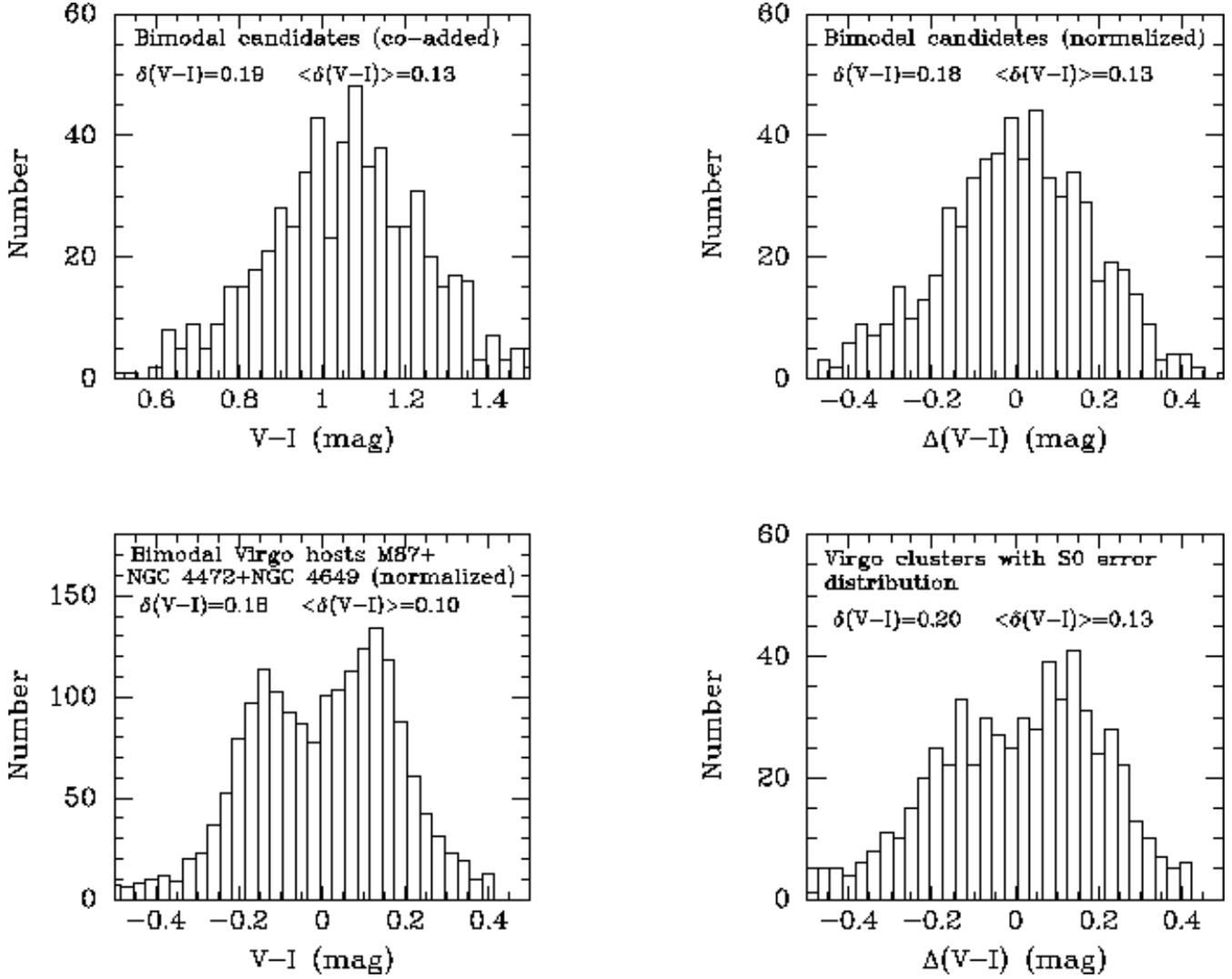}}
\caption{Top left: The co-added color distribution of the 5 galaxies with probable bimodal color distributions; Top right: The co-added color distribution of the 5 galaxies with probable bimodal color distributions normalized to a common mean color of V-I=0; Bottom left: The normalized color distribution of M87, NGC 4472 \& NGC 4649 from the analysis of Kundu et al. (1999) and Paper I show obvious bimodality; Bottom right: The subset of the normalized color distribution of the Virgo galaxies with the exact same uncertainty distribution as the bimodal S0 candidates in the upper panel. \label{fig4}}
\end{figure}

\begin{figure}
\centerline{\psfig{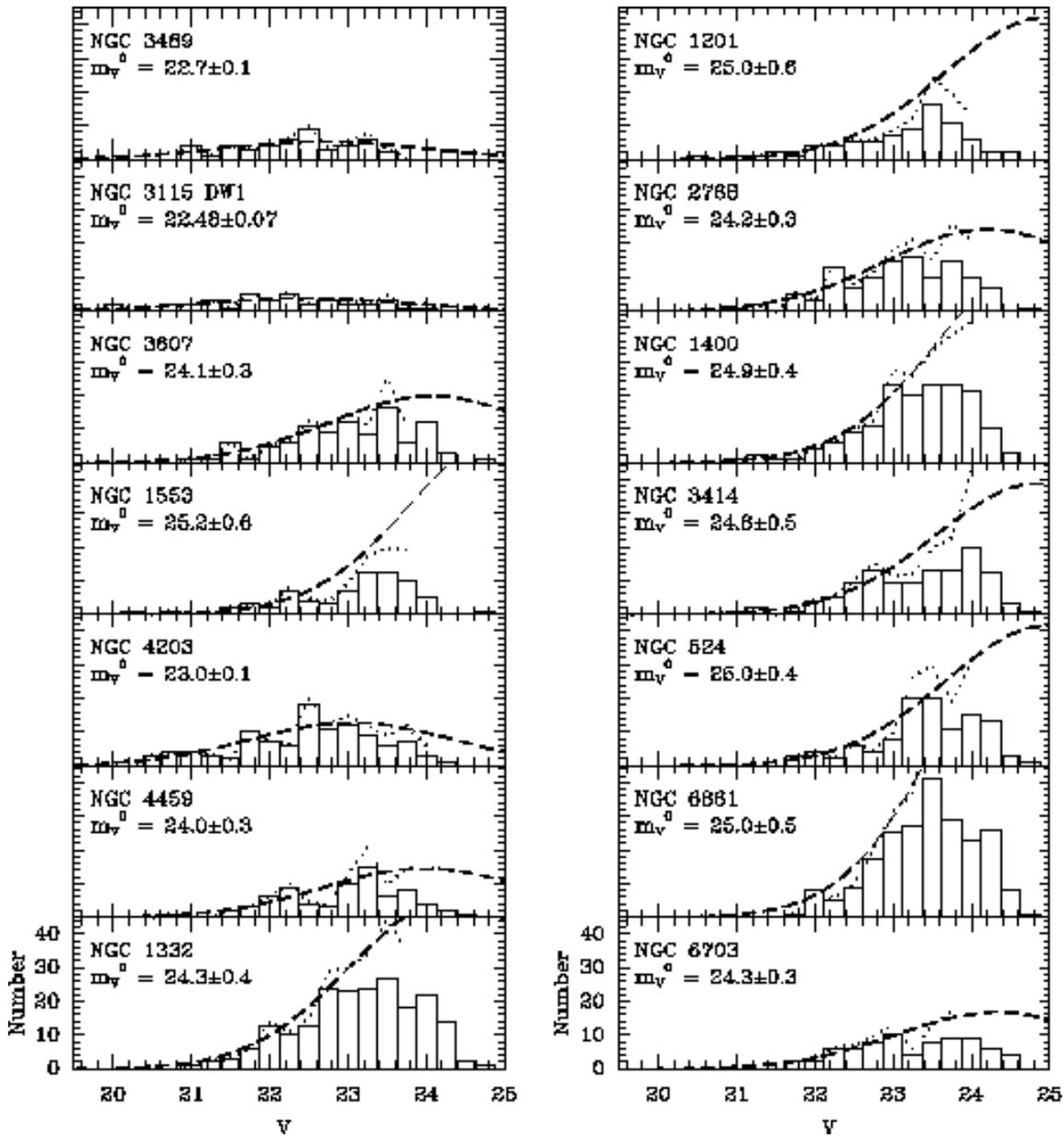}}
\caption{ a: The V-band globular cluster luminosity function   of the 14 richest cluster systems sorted by Table 1 distances. Only cluster candidates in the  range  0.5$<$V-I$<$1.5 and V$>$(m-M)-11.0 have been selected in order to minimize contaminating sources. The dotted lines trace the 
completeness corrected distribution up to the 50$\%$ completeness limit. The dashed lines mark the best fit Gaussian curve.  
b: The corresponding GCLFs in the I-band. The GCLFs are typically $\sim$1 mag brighter in I. \label{fig5}}
\end{figure}
\clearpage
\centerline{\psfig{figure=fig5b.epsi,width=6in}}
\clearpage

\begin{figure}
\centerline{\psfig{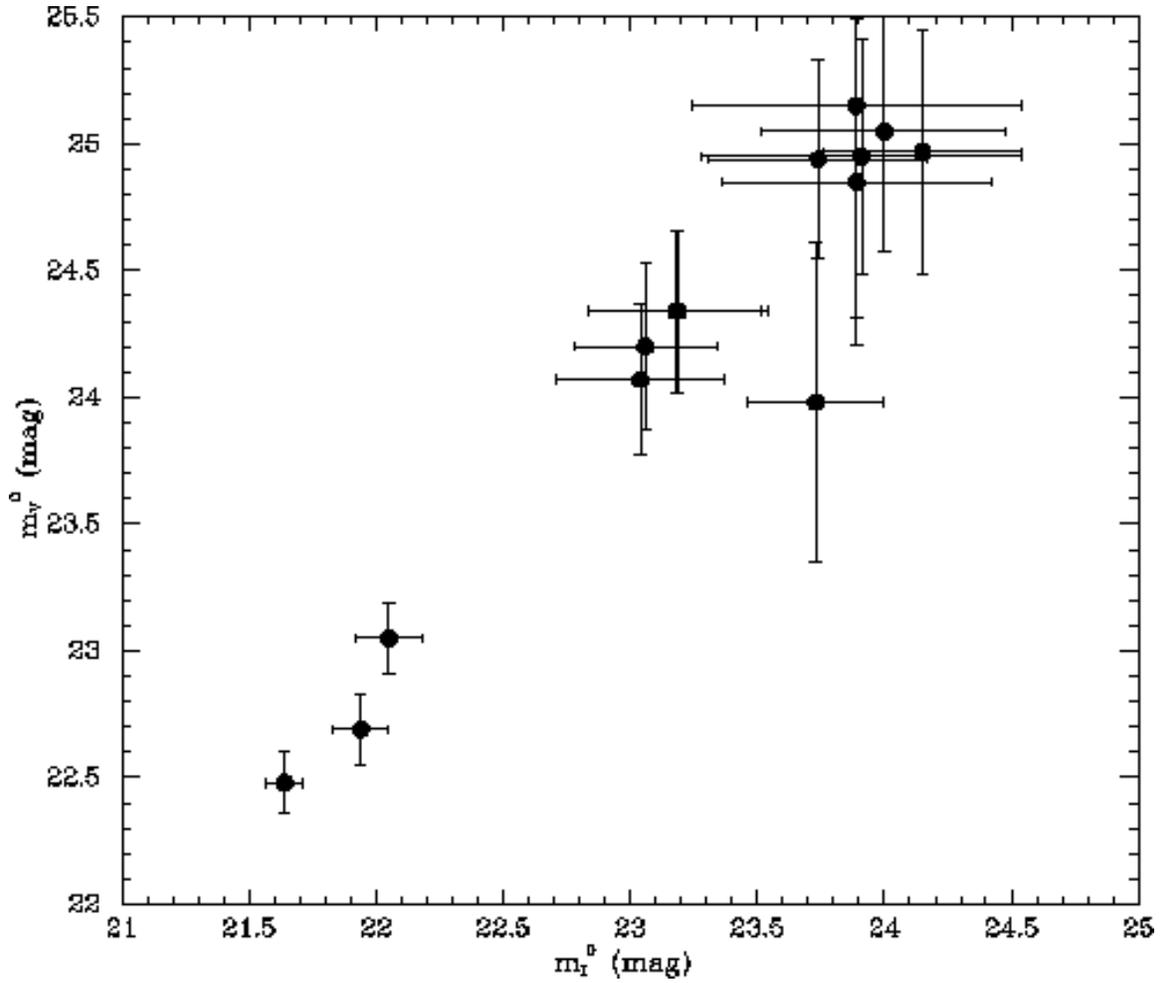}}
\caption{The turnover luminosity of the best fitting Gaussian to the GCLF in
 the V-band, m$_V^0$ vs the turnover luminosity in the I-band, m$_I^0$. The turnover luminosities in the two filters are well correlated.  
\label{fig6}}
\end{figure}

\begin{figure}
\centerline{\psfig{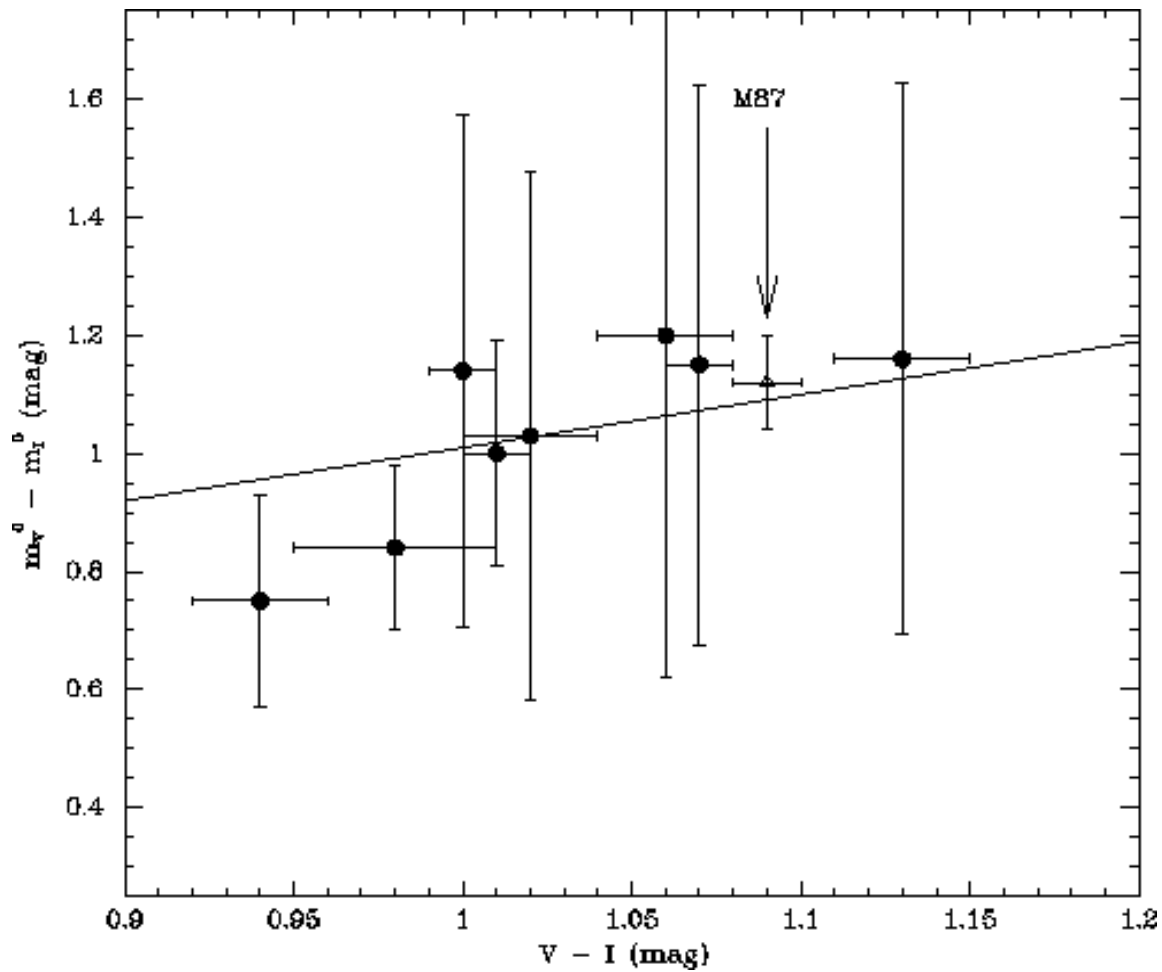}}
\caption{  The difference between V and I-band turnovers as a function of 
cluster color (metallicity).  m$_V^0$-m$_I^0$ increases with color, as predicted by the Ashman, Conti, \& Zepf (1995). The dashed line tracing the best-fit straight line from a similar analysis of ellipticals in Paper I is in good agreement with the observed trend.  
\label{fig7}}
\end{figure}

\begin{figure}
\centerline{\psfig{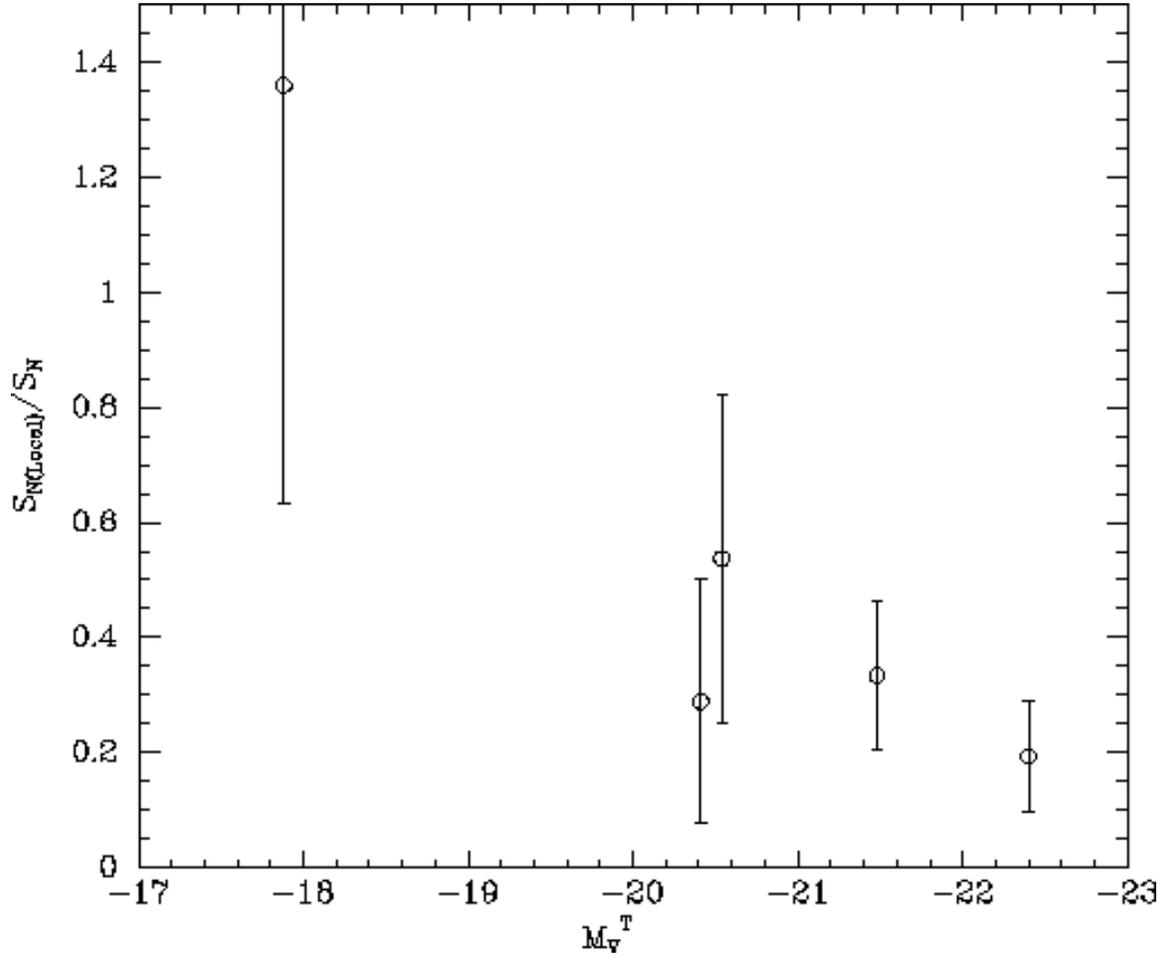}}
\caption{The ratio of S$_{N(Loc)}$/S$_N$ vs the absolute magnitude of the 
host galaxy for the galaxies which have ground based measurements of the specific frequency. The local specific frequency is generally lower than the global value with a trend of decreasing S$_{N(Loc)}$/S$_N$ with host luminosity (mass). 
\label{fig8}}
\end{figure}

\begin{figure}
\centerline{\psfig{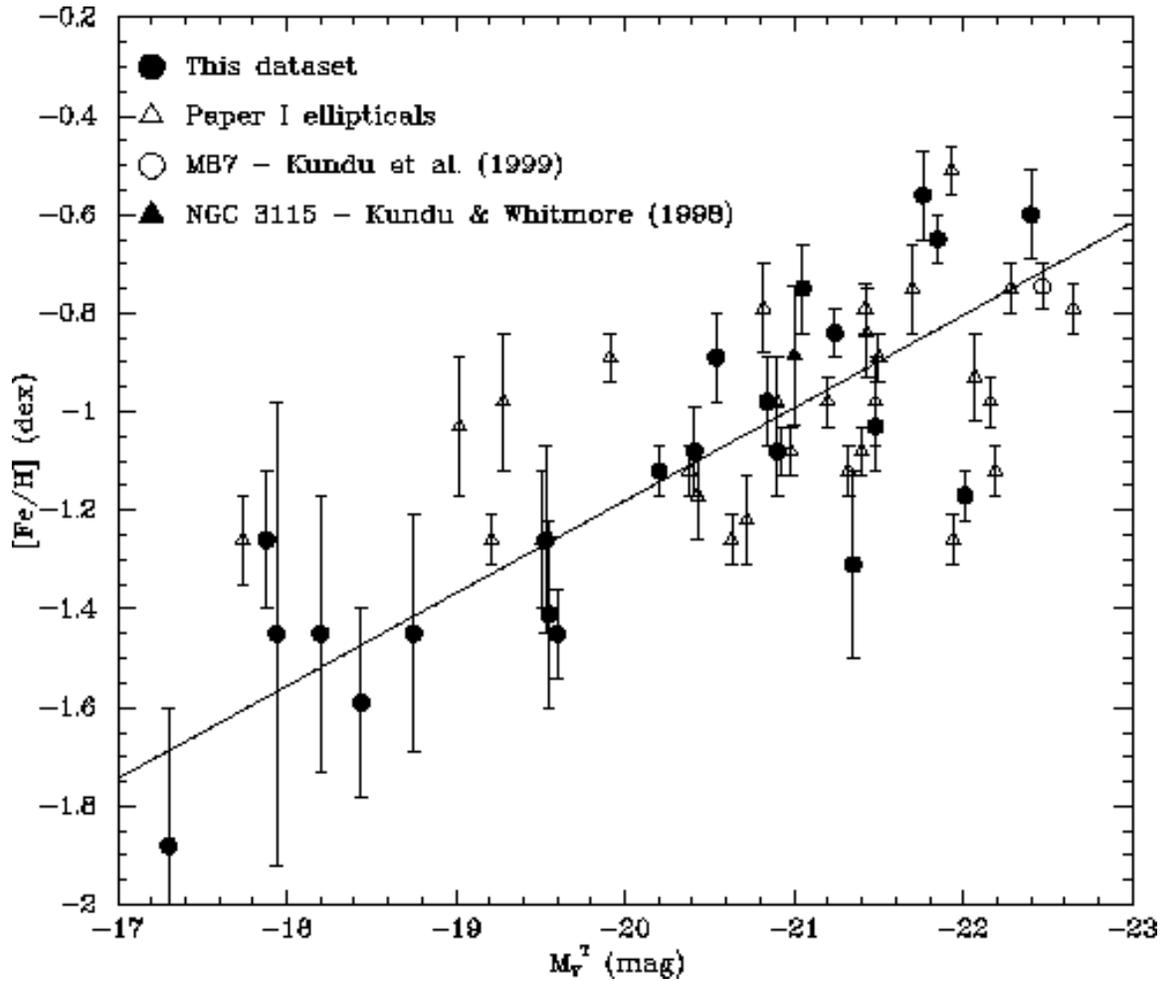}}
\caption{The average metallicity of the cluster systems  vs the absolute magnitude of the host galaxy. Only S0 systems with N$_{bg}$/N$_{cand}$$<$0.25 have been plotted as these have reliable color estimates. The elliptical galaxies from Paper 1, M87 and NGC 3115 have also been plotted for comparison.
The metallicity of the GCSs of S0s increases with host luminosity, consistent with the  trend seen for ellipticals.  The solid line 
traces the best fit straight line through the S0 data set.
\label{fig9}}
\end{figure}

 \begin{figure}
\centerline{\psfig{figure=fig10.epsi,height=7.5in,angle=0}}
\caption{ Top: The variation in the local specific frequency with mean color of the cluster system. Only systems with N$_{bg}$/N$_{cand}$$<$0.25 have been plotted as these have reliable color estimates. Elliptical galaxies from Paper I have also been shown. Bottom: The variation of the local specific frequency within our field of view with
host galaxy brightness for the S0 sample from this paper and the elliptical galaxies from Paper I. Only elliptical candidates with $\delta$S$_N$ $<$ 3 and S0s with  $\delta$S$_N$ $<$ 1.5 are plotted. As explained in the text the uncertainty in the local specific frequency of S0s is underestimated by a factor of 2 as compared to the ellipticals.
\label{fig10}}
\end{figure}

\begin{figure}
\centerline{\psfig{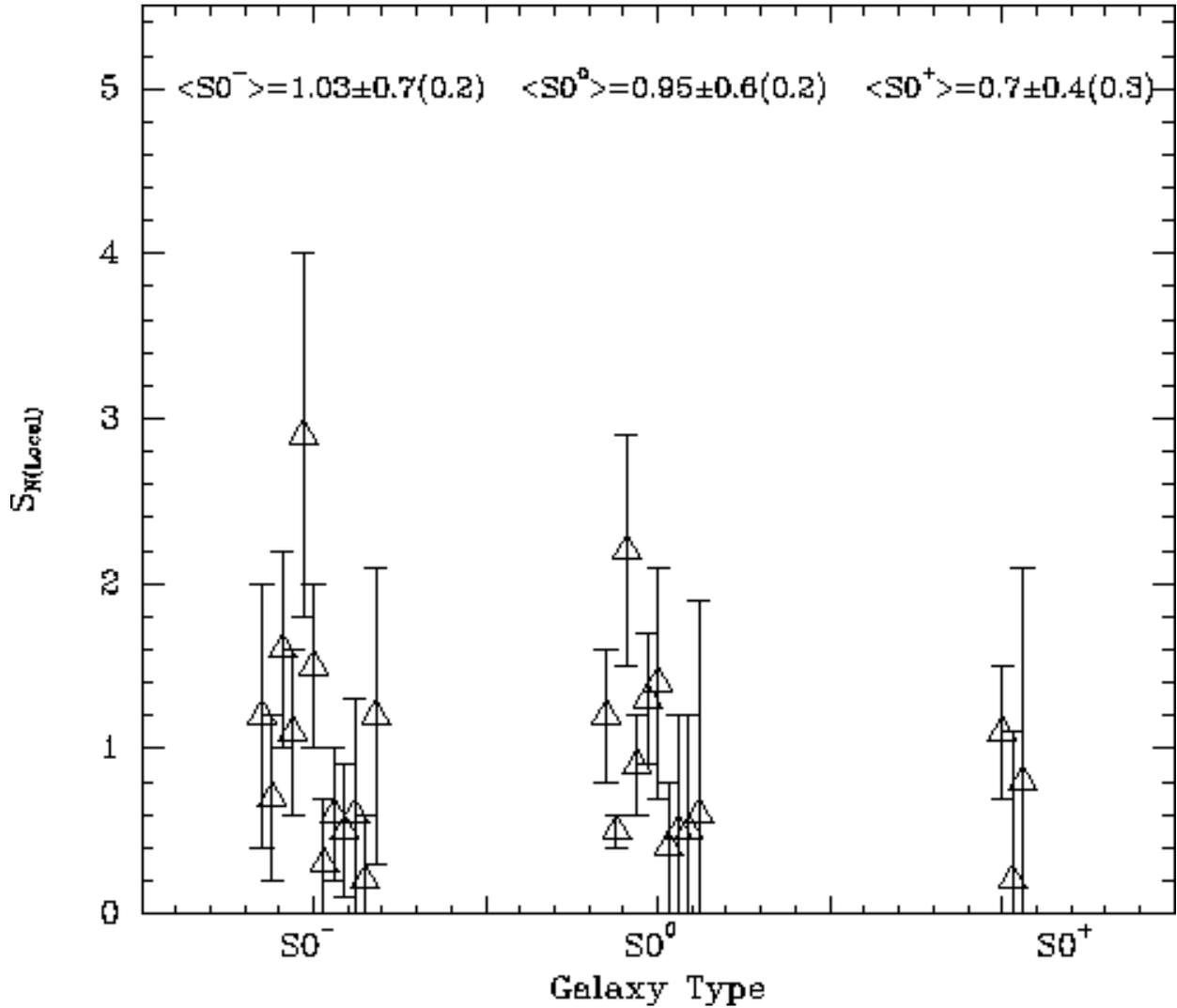}}
\caption{ The variation of the local specific frequency with Hubble type
(see text for details of our binning parameters). The positions of the points
have been shifted by arbitrary small amounts along the abscissa to
 help distinguish individual data points. Only  candidates with $\delta$S$_N$ $<$ 1.5 are plotted. There is  a weak trend of later type S0s having lower specific frequencies.
\label{fig11}}
\end{figure}

\begin{figure}
\centerline{\psfig{figure=fig12.epsi,height=7.5in,angle=0}}
\caption{Top: The local specific frequencies of globular 
cluster systems in S0s and ellipticals (from Paper I) as a function of the log of the number of galaxies in the host galaxy cluster. The selection criteria for plotting a galaxy are the same as in Fig 10.
 Bottom: The average metallicity of the GCSs in ellipticals and S0s vs the log of the number of galaxies in the host galaxy cluster.
\label{fig12}
}
\end{figure}

\begin{figure}
\centerline{\psfig{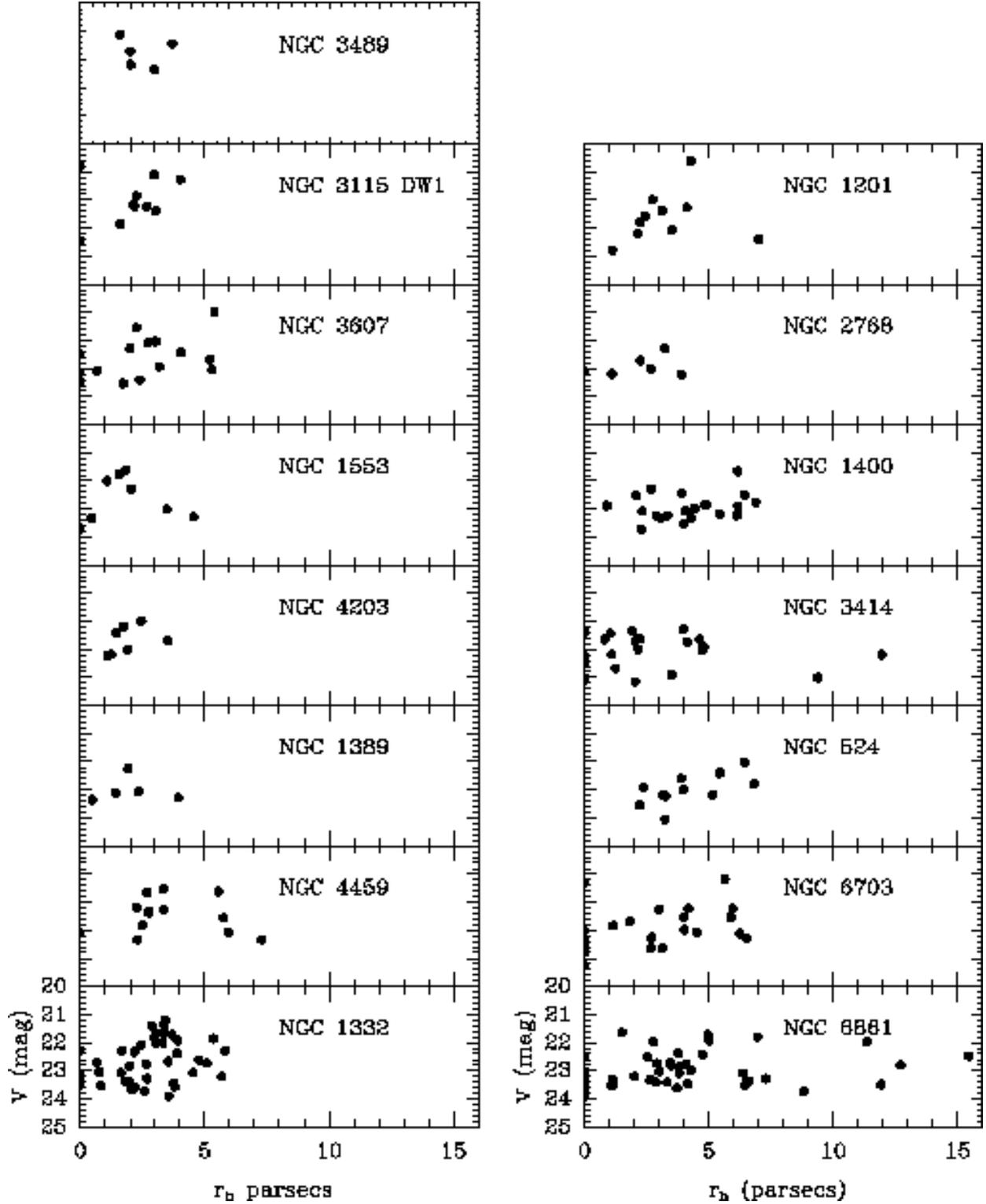}}
\caption{The half-light radius vs V-band magnitude of globular clusters in the PC. Only galaxies with at least 5 cluster candidates in the PC are plotted.
The half light radii of clusters in all the galaxies largely fall in a narrow range of sizes between 0-7 pc. The half light radius does not appear to depend on the luminosity (mass) of the clusters.
\label{fig13}
}
\end{figure}

\begin{deluxetable}{lllcllll}
\large

\tablenum{1}
\tablecaption{General Properties of the Program Galaxies}
\startdata
\\
\tableline
\tableline
Galaxy 	       & Morph. Typ		        & Gp/Clstr\tablenotemark{a}    & Hel. Vel & Dist\tablenotemark{b} &    A$_V$\tablenotemark{c} & M$_V^{T^0}$\tablenotemark{d}   \\
	       &		 	  	     &             & Km s$^{-1}$ & Mpc & mag  & mag       \\
 (1)	       & (2)	       		  & (3)	     & (4)	   & (5)   & (6)      & (7)	       \\
\tableline
NGC 524        &  SA(rs)0$^+$            & NGC 524Gp   & 2421  & 32.6     & 0.10      &	 -22.4	  \\
NGC 2768       & S0$_{1/2}$/E6,LIN	 & NGC 2768Gp  & 1339  & 22.8     & 0.12      &	 -22.0	  \\
NGC 6861       & SA(s)0$^-$	         & Telescopium & 2819  & 36.6     &  0.11     &	 -21.8	  \\
		 			
NGC 6703       & SA0$^-$		 &     	   & 2365  & 36.6     & 0.20      &	 -21.8 	  \\
NGC 1553       & SA(rl)0$^0$,LIN         & Dorado      & 1080  & 14.9     & 0.0       &	 -21.5	  \\
NGC 474        & (R$'$)SA(s)0$^0$        & NGC 488Gp   & 2372  & 31.6     & 0.04      &	 -21.4	  \\
							 
NGC 1332       &  S(s)0$^-$:sp           & Eridanus    & 1524  & 20.0     & 0.02      &	 -21.2	  \\
NGC 3414       & S0 pec	                 & LGG 227     & 1414  & 24.5     & 0.0       &	 -21.0	  \\
NGC 4459       & SA(r)0$^+$	         & Virgo I	   & 1210  & 17.3\dag & 0.06      &	 -20.9	  \\
							 
NGC 1201       &  SA(r)0$^0$             & NGC 1255Gp  & 1671  & 20.5     & 0.0       &	 -20.8	  \\
NGC 1400       &  SA0$^-$                & Eridanus A  & 558   & 22.9\dag & 0.10      &	 -20.6	  \\
NGC 3607       & SA(s)0$^0$              & Leo II      & 935   & 11.4     & 0.0       &	 -20.4	  \\
							 
NGC 4203       & SAB0$^-$,LIN	         & Coma I 	   & 1086  & 15.0     & 0.0       &	 -20.2	  \\
NGC 2902       & SA(s)0$^0$              & LGG 174     & 1990  & 27.4     & 0.19      &	 -20.2	  \\
NGC 3489       & SAB(rs)0$^+$	         & Leo I       &  708  & 9.7      & 0.02      &	 -19.6	  \\
							 
NGC 4379       & S0$^-$ pec:	         & Virgo I	   & 1069  & 17.3\dag & 0.02      &	 -19.6	  \\
NGC 1389       &  SAB(s)0$^-$            & Fornax      & 986   & 16.4     & 0.0       &	 -19.5	  \\
NGC 3056       & SA(rl)0$^0$             &             & 1017  & 12.4     & 0.12      &	 -18.9	  \\
							 
NGC 3156       & S0	                 & NGC 3166Gp  & 1118  & 16.3     & 0.03      &	 -18.9	  \\
IC 3131        & d:S0$_1$(0),N:	         &Virgo W$'$?  &1596   & 30.0\dag & 0.0       & -18.9\ddag	  \\
NGC 1375       &  SAB0$^0$:sp            & Fornax      & 740   & 16.4     & 0.0       &	 -18.8	  \\
							 
VCC 165        & S0$_1$(0,2):	         & Virgo M?    & 255   & 30.0\dag & 0.02      & -18.5\ddag	  \\
NGC 3599       & SA0	                 & Leo II      & 781   & 11.4     & 0.0       &	 -18.4	  \\
NGC 2328       & (R$'$)SAB0$^-$          &             & 1159  & 13.5     & 0.33      &	 -18.4	  \\
							 
NGC 4431       & SA(r)0	                 & Virgo I     & 913   & 17.3\dag & 0.06      &	 -18.4	  \\
IC 1919        &  SA(rs)0$^-$?           & Fornax      & 1220  & 16.4     & 0.0       & -18.2\ddag	  \\
NGC 1581       &  S0$^-$                 & Dorado      & 1600  & 14.9     & 0.0       &	 -18.2	  \\
							 
ESO 358- G 059 &  SAB0$^-$               & Fornax      & 1007  & 16.4     & 0.0       &	 -18.0	  \\
NGC 3870       & S0?		         & U Ma NED3   & 756   & 15.5     & 0.0       &	 -17.9	  \\
NGC 3115 DW1   & SA(s)0$^0$ pec          &             & 715   & 11.0     & 0.13      &	 -17.9	  \\
							 
ESO 118- G 034 &  S0$^0$ pec             & NGC 1672Gp  & 1178  & 13.3     & 0.0       &	 -17.6	  \\
IC 3540        & SB0	                 & Virgo I     & 753   & 17.3\dag & 0.08      & -17.3\ddag	  \\
NGC 4150       & SA(r)0$^0$?             & C Vn I      & 226   & 4.3\dag  & 0.03      &	 -16.6	  \\
							 
NGC 404        & SA(s)0$^-$              & LGG 11      & -48   & 3\dag    & 0.17      &	 -16.5	  \\

\enddata
\tablenotetext{} {1 Galaxy; 2 Morphological classification; 3 Group/Cluster membership; 4 Heliocentric velocity; 5 Distance using velocities; 6 Reddening in V; 7 V-band absolute magnitude }
\tablenotetext{a} {From Garcia (1993), Huchra \& Geller (1982), Geller \& Huchra
 (1983) and Fouqu\'e {\it et al.} (1992)}
\tablenotetext{b} { Using the Virgocentric infall model from Postman \& Lauer (1995) and 
assuming H$_0$ = 75 Km s$^{-1}$ Mpc$^{-1}$. See notes below for details on the 
ones marked with \dag.}
\tablenotetext{c} { Using A$_B$ from the NED extragalactic database and the 
extinction law from Mathis (1990).  }
\tablenotetext{d} { Using m$_V^{T^0}$ from the RC3 catalog and the distances in 
column 6. See notes below for the ones marked with \ddag }
\tablecomments {\dag Galaxies in  the Virgo cluster (VCC 165, IC 3131, NGC 4379,
 NGC 4431, NGC 4459, IC3540) return three possible solutions to the equation of 
motion. The most likely sub-cluster membership has been assigned based on their 
spatial position from the Virgo cluster maps of Bingelli, Tammann \& Sandage
 (1985).  Virgo I is assumed to have a recessional velocity of 1300 
 Km s$^{-1}$ the M, W and W$'$ clouds are assumed to be roughly twice as far 
(Bingelli et al. 1985). Although NGC 1400 appears to lie
in the Eridanus cluster it has an anomalously low heliocentric velocity of
558 Km s$^{-1}$. Surface brightness fluctuation measurements by Jensen,  Tonry
 \& Luppino (1998) confirm that NGC 1400 is indeed a  part of the Eridanus cluster,
hence we adopt their distance of 22.9 Mpc to Eridanus. Though the 
negative radial velocity of NGC 404 suggests that it is a member of the local
group, Wilkind \& Henkel (1990) claim that it is at a distance of 10 Mpc. We argue in 
$\S$3 that the distance to NGC 404 is most likely to be $\approx$3 Mpc and we adopt
this estimate for our analysis. NGC 4150 appears to be member of the nearby
 Canes Venatici cloud, most of whose members are 2-8 Mpc from the Milky Way . We
 adopt the average distance  of 4.3 Mpc to C Vn (Makarova {\it et al.}
 1998) with the cautionary note that NGC 4150 is located near the Southern
edge of the cloud and the candidate  member galaxy nearest to it, 
UGC 7131, appears to be a background galaxy that is further than 14 Mpc  according to
 Makarova {\it et al.} (1998). \\[0.1cm] \ddag  IC 1919 \& IC 3540 have only B band 
measurements of their total magnitude. B-V $\approx$ 0.9, typical of S0 galaxies, 
was used to determine the V magnitudes. Since no measurement of the surface brightness
 VCC 165 \& IC 3131 exist  in the literature we estimate m$_V^{T^0}$ from our data.
 }
\end{deluxetable}

\begin{deluxetable}{cllllllll}
\large
\tablenum{2}
\tablecaption{Globular Cluster Properties}
\startdata
\\
\tableline
\tableline
Galaxy         & $<$V-I$>$     &  [Fe/H] 	& N$_{cand}$ & N$_{bg}$ & N$_{Tot}$ &  M$_V^{FOV}$	& S$_{N(Local)}$ & S$_N$$^\ddag$ \\
	       & mag           & dex	   	&	     &	&	& mag	& \\
(1)	       &   (2)	       &   (3)		&  (4)	      &   (5)	  &  (6)	 &    (7)	&   (8)		 & (9)	 \\	
\tableline
NGC  524 &   1.12 (0.02) &   -0.6 (0.09) &  113 &  9 &     645 &      -21.9 &      1.1$\pm$0.4  & 5.7$\pm$1.8 \\
NGC 2768 &     1. (0.01) &  -1.17 (0.05) &  118 &  6 &     343 &      -21.2 &      1.2$\pm$0.4 \\
NGC 6861 &   1.11 (0.01) &  -0.65 (0.05) &  235 & 25 &    1858 &      -21.8 &      3.6$\pm$1.6 \\
NGC 6703 &   1.13 (0.02) &  -0.56 (0.09) &   82 & 16 &     560 &      -21.7 &      1.2$\pm$0.8 \\
NGC 1553 &   1.03 (0.02) &  -1.03 (0.09) &   70 &  7 &     114 &       -21. &      0.5$\pm$0.1  & 1.5$\pm$0.5 \\
NGC  474 &   0.97 (0.04) &  -1.31 (0.19) &   34 &  2 &     185 &       -21. &      0.7$\pm$0.5 \\
NGC 1332 &   1.07 (0.01) &  -0.84 (0.05) &  204 & 10 &     501 &      -20.9 &      2.2$\pm$0.7 \\
NGC 3414 &   1.09 (0.02) &  -0.75 (0.09) &  117 &  4 &     397 &       -21. &      1.6$\pm$0.6 \\
NGC 4459 &   1.02 (0.02) &  -1.08 (0.09) &   75 &  3 &     148 &      -20.6 &      0.9$\pm$0.3 \\
NGC 1201 &   1.04 (0.02) &  -0.98 (0.09) &   77 &  9 &     177 &      -20.5 &      1.1$\pm$0.5 \\
NGC 1400 &   1.06 (0.02) &  -0.89 (0.09) &  158 &  3 &     419 &      -20.4 &      2.9$\pm$1.1  & 5.4$\pm$2 \\
NGC 3607 &   1.02 (0.02) &  -1.08 (0.09) &   98 &  6 &     130 &       -20. &      1.3$\pm$0.4  & 4.5$\pm$3 \\
NGC 4203 &   1.01 (0.01) &  -1.12 (0.05) &  106 &  5 &     175 &      -20.2 &      1.5$\pm$0.5 \\
NGC 2902 &   1.17 (0.04) &  -0.37 (0.19) &   15 &  5 &      35 &      -20.2 &      0.3$\pm$0.4 \\
NGC 3489 &   0.94 (0.02) &  -1.45 (0.09) &   46 &  4 &      51 &      -18.9 &      1.4$\pm$0.7 \\
NGC 4379 &   0.95 (0.04) &  -1.41 (0.19) &   22 &  3 &      37 &      -19.6 &      0.6$\pm$0.4 \\
NGC 1389 &   0.98 (0.04) &  -1.26 (0.19) &   19 &  2 &      32 &      -19.5 &      0.5$\pm$0.4 \\
NGC 3056 &   1.07 (0.04) &  -0.84 (0.19) &   25 &  8 &      23 &      -18.9 &      0.6$\pm$0.7 \\
NGC 3156 &   1.02 (0.09) &  -1.08 (0.42) &   11 &  3 &      13 &      -18.9 &      0.4$\pm$0.4 \\
IC 3131  &                 &                 &    3 &  1 &       8 &      -18.9 &      0.2$\pm$0.4 \\
NGC 1375 &   0.94 (0.05) &  -1.45 (0.24) &   11 &  1 &      17 &      -18.7 &      0.5$\pm$0.7 \\
VCC  165 &                 &                 &   11 &  6 &       0 &      -18.5 &       0.$\pm$2.1 \\
NGC 3599 &   0.91 (0.04) &  -1.59 (0.19) &   21 &  1 &      29 &      -18.4 &      1.2$\pm$0.9 \\
NGC 2328 &                 &                 &   18 & 13 &       0 &      -18.4 &       0.$\pm$1.4 \\
NGC 4431 &   0.92 (0.06) &  -1.55 (0.28) &   10 &  2 &      11 &      -18.4 &      0.5$\pm$0.7 \\
IC 1919  &   0.94 (0.06) &  -1.45 (0.28) &   13 &  2 &      21 &      -18.2 &      1.1$\pm$1.7 \\
NGC 1581 &   1.14 (0.08) &  -0.51 (0.38) &    8 &  5 &       4 &      -18.2 &      0.2$\pm$0.9 \\
ESO 358-G059& 1.07 (0.09)&  -0.84 (0.42) &    7 &  2 &       9 &      -17.9 &      0.6$\pm$1.3 \\
NGC 3870 &   0.94 (0.1)  &  -1.45 (0.47) &    8 &  1 &      13 &      -17.9 &      0.8$\pm$1.3 \\
NGC 3115 DW1& 0.98 (0.03)&  -1.26 (0.14) &   37 &  5 &      42 &       -17. &      6.8$\pm$2.4  & 5$\pm$2 \\
ESO 118-G034&              &                 &    9 &  6 &       3 &      -17.6 &      0.3$\pm$2.  \\
IC 3540  &   0.85 (0.06) &  -1.88 (0.28) &   14 &  1 &      26 &      -17.3 &      3.2$\pm$4.2 \\
NGC 4150 &   0.92 (0.06) &  -1.55 (0.28) &   10 &  2 &       7 &      -16.6 &      1.7$\pm$1.8 \\
NGC  404 &		   &		     &   83 & 48					   \\
\tableline		       
Avg$^b$	 & 1.00$\pm$0.07$^a$   & -1.1$\pm$0.3$^a$ &	 &      &	&	  &  1.0$\pm$0.6$^\dag$  \\
\enddata
\tablenotetext{} {1 Galaxy Name; 2 V-I color in mag; 3 [Fe/H] in dex; 4 No of candidate objects in the color range
 0.5$<$V-I$<$1.5 ; 5 No of objects with colors in the the range 0.0$<$V-I$<$0.5
 or 1.5$<$V-I$<$2.0; 6 Calculated total number of globular clusters in the 
field of view; 7 Total V band luminosity of the galaxy within the field of view; 8 The local Specific Frequency
of globular clusters; 9 Global Specific Frequency from the compilation of Kissler-Patig (1997) and references therein. }
\tablenotetext{\dag} { Avg S$_{N(Loc)}$ of systems with $\delta$S$_{N(Loc)}$ $<$ 1.5   }
\tablenotetext{\ddag} { The values of S$_N$ have been adjusted to Table 1 distances.   }
\tablenotetext{a} { Only systems with N$_{bg}$/N$_{cand}$ $<$ 0.25 have been considered in the calculation of the mean colors and metallicities. }
\tablenotetext{b} {The average color, metallicity and specific frequency, and 
associated uncertainties, are unchanged by the addition of the NGC 4550 values from Paper 1.}

\end{deluxetable}

\begin{deluxetable}{llllllll}
\tablenum{3}
\tablecaption{Distances from the GCLF}
\startdata
\\
\tableline
\tableline
Galaxy & m$_V^0$ & m$_I^0$ & (m-M)$_{Vel}$ & (m-M)$_{Lit}$  & (m-M)$_V$$^\dag$   & (m-M)$_I$$^\dag$  & $<$m-M$>_{GCLF}$$^\dag$ \\      
 (1) & (2) & (3) & (4) & (5) & (6) & (7) & (8) \\
\tableline
NGC 3489    & 22.69$\pm$0.11 &  21.94$\pm$0.14 &  29.9 &    30.  &  30.1$\pm$0.1  &     30.4$\pm$0.1  &     30.25$\pm$0.09 \\
NGC 3115 DW1& 22.48$\pm$0.07 &  21.64$\pm$0.12 &  30.2 &         & 29.89$\pm$0.08  &     30.1$\pm$0.1  &     30.$\pm$0.07 \\
NGC 3607    & 24.07$\pm$0.33 &  23.04$\pm$0.3 &   30.3 &   31.5  & 31.48$\pm$0.3  &     31.5$\pm$0.3   &     31.49$\pm$0.22 \\
NGC 1553    & 25.15$\pm$0.65 &  23.89$\pm$0.84 &  30.9 &   30.8  &                 &                    &                    \\
NGC 4203    & 23.05$\pm$0.13 &  22.05$\pm$0.14 &  30.9 &         & 30.46$\pm$0.1  &    30.51$\pm$0.1  &     30.49$\pm$0.10 \\
NGC 4459    & 23.98$\pm$0.27 &  23.73$\pm$0.63 &  31.2 &   30.9  &                 &                    &                    \\
NGC 1400    & 24.94$\pm$0.43 &  23.74$\pm$0.39 &  31.5 &   31.9  & 32.35$\pm$0.4  &     32.2$\pm$0.4  &     32.28$\pm$0.29 \\
NGC 1332    & 24.34$\pm$0.35 &  23.19$\pm$0.32 &  31.5 &   31.5  & 31.75$\pm$0.4  &    31.65$\pm$0.3  &     31.7$\pm$0.24 \\
NGC 1201    & 24.95$\pm$0.63 &  23.91$\pm$0.46 &  31.6 &         &                 &                    &                    \\
NGC 2768    & 24.2$\pm$0.28 &  23.06$\pm$0.33 &    31.8 &   31.6  & 31.61$\pm$0.3  &    31.52$\pm$0.3 &     31.57$\pm$0.22 \\
NGC 3414    & 24.85$\pm$0.53 &  23.89$\pm$0.64 &  31.9 &   31.8  &                 &                    &                    \\
NGC  524    & 24.97$\pm$0.39 &  24.15$\pm$0.48 &  32.6 &         &                 &                    &                    \\
NGC 6703    & 24.34$\pm$0.34 &  23.18$\pm$0.32 &  32.8 &   32.4  &  31.75$\pm$0.3 &     31.64$\pm$0.3 &     31.70$\pm$0.23  \\
NGC 6861    & 25.05$\pm$0.48 &    24.$\pm$0.47 &  32.8 &         &                 &                    &                    \\
\\
\tableline
\enddata
\tablenotetext{} {1 Galaxy Name; 2 GCLF V-band turnover; 3 GCLF I-band turnover; 4 Distance modulus from recessional velocity; 5 Distance modulus from 
the literature. See text for details ; 6 Distance modulus from V Band GCLF; 7 Distance modulus from I band GCLF; 
8 Mean distance modulus from V and I band GCLFs  }
\tablenotetext{\dag} { GCLF distances are reported only for those galaxies with uncertainty in the mean GCLF distance less than 0.3 mag.   }
\end{deluxetable}

\begin{deluxetable}{llll}
\tablenum{4}
\tablecaption{Cluster Sizes in the PC }
\startdata
\\
\tableline
\tableline
Galaxy   &  Number & r$_h$   \\
	 &  in PC & $pc$    \\
1	 & 2        &  3     \\
\tableline
 NGC 3489  &   5     &  2.0  \\
 NGC 3115 DW1 &  10  &  2.2  \\
 NGC 3607  &  15     &  2.4  \\
 NGC 1553  &   8     &  1.7  \\
 NGC 4203  &   8     &  1.7  \\
 NGC 1389  &   5     &  1.9  \\
 NGC 4459  &  13     &  2.8  \\
 NGC 1332  &  41     &  3.0  \\
 NGC 1201  &  10     &  2.9  \\
 NGC 2768  &   6     &  2.5  \\
 NGC 1400  &  21     &  4.0  \\
 NGC 3414  &  24     &  2.0  \\
 NGC  524  &  11     &  3.9  \\
 NGC 6703  &  23     &  2.7  \\
 NGC 6861  &  46     &  3.4  \\
\tableline
Avg &	&  2.6$\pm$0.7 (0.26) \\

\enddata
\tablenotetext{} {1 Galaxy; 2 Number of cluster candidates in the PC chip; 3 Median half-light radius in parsecs. }
\\
\end{deluxetable}

\end{document}